\documentclass[conference]{IEEEtran}
\IEEEoverridecommandlockouts
\usepackage{cite}
\usepackage{flushend}

\usepackage{amsmath,amssymb,amsfonts}
\usepackage{algorithm,algpseudocodex}
\usepackage{graphicx}
\usepackage{textcomp}
\usepackage{booktabs}
\usepackage{subcaption}
\usepackage{listings}
\usepackage{url}
\usepackage{hyperref}
\usepackage{xcolor}

\usepackage[margin=1in]{geometry}
\usepackage{soul} 

\def\BibTeX{{\rm B\kern-.05em{\sc i\kern-.025em b}\kern-.08em
    T\kern-.1667em\lower.7ex\hbox{E}\kern-.125emX}}
\begin{document}
\title{
GPU-Accelerated Modified Bessel Function of the Second Kind for Gaussian Processes}

 \author{
 \IEEEauthorblockN{Zipei Geng$^1$, Sameh Abdulah$^2$, Ying Sun$^1$, Hatem Ltaief$^2$, David E. Keyes$^2$, Marc G. Genton$^1$}
 \IEEEauthorblockA{
 $^1$Statistics Program, \\
 $^2$Applied Mathematics and Computational Sciences (AMCS) Program, \\
 $^{1,2}$King Abdullah University of Science and Technology, Thuwal, Saudi Arabia. \\
 $^{1,2}$\{firstname.lastname\}@kaust.edu.sa
 }
}

\maketitle

\begin{abstract}
Modified Bessel functions of the second kind are widely used in physics, engineering, spatial statistics, and machine learning. Since contemporary scientific applications, including machine learning, rely on GPUs for acceleration, providing robust GPU-hosted implementations of special functions, such as the modified Bessel function, is crucial for performance. Existing implementations of the modified Bessel function of the second kind rely on CPUs and have limited coverage of the full range of values needed in some applications. In this work, we present a robust implementation of the modified Bessel function of the second kind on GPUs, eliminating the dependence on the CPU host. We cover a range of values commonly used in real applications, providing high accuracy compared to common libraries like the GNU Scientific Library (GSL) when referenced to Mathematica as the authority. Our GPU-accelerated approach also demonstrates a $2.68$X performance improvement using a single A100 GPU compared to the GSL on 40-core Intel Cascade Lake CPUs.
Our implementation is integrated into \emph{ExaGeoStat}, the HPC framework for Gaussian process modeling, where the modified Bessel function of the second kind is required by the Mat\'ern covariance function in generating covariance matrices. We accelerate the matrix generation process in \emph{ExaGeoStat} by up to $12.62$X with four A100 GPUs while maintaining almost the same accuracy for modeling and prediction operations using synthetic and real datasets.
\end{abstract}


\section{Introduction}
The modified Bessel function of the second kind, denoted as \( K_\nu(x) \), where \( \nu \) represents the order and \( x \) the argument, arises in a wide range of applications across mathematics, physics, engineering, spatial statistics, and machine learning. For instance, in physics and engineering, \( K_\nu(x) \) frequently appears in solving PDEs in cylindrical symmetry with radially decaying behavior. Examples include heat conduction, damped wave propagation~\cite{sharma2007damped}, electromagnetic fields in cylindrical waveguides~\cite{karamian2022role}, and Schr\"{o}dinger's equation in axisymmetry~\cite{martin2021quasi}. 
Similarly, in Gaussian processes, \( K_\nu(x) \) is used in the Mat\'{e}rn kernel function, which is widely used to construct correlation matrices that capture spatial relationships between locations~\cite{wang2023parameterization}. These diverse applications emphasize the importance of efficient computation of the \( K_\nu(x) \) function.

In the literature, three methods are widely employed to compute the modified Bessel function of the second kind across a broad range of values for \(x\) and \(\nu\)~\cite{bbtakekawa}: series expansions~\cite{bbtemme}, continued fractions~\cite{amos1974computation}, and asymptotic expansions~\cite{olver2009bessel}.  The first expresses \( K_\nu(x) \) as an infinite series that involves powers of \(x\). This approach is particularly effective for small \(x\), where truncating the series to a finite number of terms provides a highly accurate approximation. 
The continued fractions method represents the Bessel function as an infinite recursive fraction. This method is suitable for larger \(x\) than is the series expansion, but it remains computationally intensive. An asymptotic expansion approximates the Bessel function by capturing its dominant behavior as \(x\) becomes very large. By neglecting lower-order terms, this method efficiently describes the primary growth trend. However, it lacks accuracy for small \(x\), where such approximations fail to converge effectively.  Combining these three methods and other less common approaches can effectively cover a wider range of \(x\) and \(\nu\) values. Nevertheless, these techniques are computationally demanding and inherently sequential, posing challenges for parallelization on modern architectures, such as GPUs.

In this paper, we adopt an integration-based method proposed by~\cite{bbtakekawa} to compute \(K_{\nu}(x)\) for real order \(\nu \in \mathbb{R}\), denoted by \textsc{BesselK}, and optimize it for GPU implementation. This approach, implemented in CUDA, is well suited for parallel execution, as it represents the \textsc{BesselK} function using quadrature over finite intervals. The key advantages of this method, beyond its parallelizability, are its flexibility and its ability to efficiently cover a reasonable range of \(x\) and \(\nu\) values.  
Our implementation has been integrated into \emph{ExaGeoStat}~\cite{abdulah2018exageostat}, a scalable geospatial data modeling and prediction framework optimized for manycore systems, including modern GPUs.   \emph{ExaGeoStat} possesses three functionalities: synthetic data generation, modeling, and prediction. Each of these operations requires, at a minimum, the generation of a covariance matrix of size \(N \times N\), where \(N\) represents the number of spatial locations in its simplest form. The Mat\'{e}rn kernel (or covariance function), which involves the \textsc{BesselK} function, is often used in Gaussian processes to construct this covariance matrix. 

Our contributions are summarized as follows:

\begin{itemize}
    \item We present a highly efficient GPU implementation of the \textsc{BesselK} function, which addresses a critical performance bottleneck encountered in numerous scientific applications.  
    
    \item {\color{black} We accelerate an integration-based algorithm using a unified integral bound for \textsc{BesselK}, enabling efficient handling of a practical range of \(x\) and \(\nu\) commonly encountered in Gaussian processes on the GPU.}

    \item We combine the accelerated integration-based algorithm with a series expansion to accurately compute the \textsc{BesselK} function for \(x < 0.1\), where the integral-based approach specifically underperforms, making our solution more comprehensive.  

    \item { \color{black} Our refined algorithm combines series expansion and integration, both of which are established methods to approximate BesselK. However, to the best of our knowledge, we are the first to deploy these approaches on massively parallel GPU hardware accelerators.}

    \item We integrate the refined algorithm into the covariance matrix generation step of \emph{ExaGeoStat} and evaluate its performance and accuracy against existing CPU-based implementations, such as the GNU Scientific Library.
    
    \item We assess the accuracy of the proposed algorithm to compute the \textsc{BesselK} function against existing approaches. Additionally, we evaluate its accuracy and overall impact within the \emph{ExaGeoStat} software pipeline using synthetic datasets and a real-world wind speed dataset for climate and weather modeling applications.  
\end{itemize}

\section{Related Work}


The modified Bessel function of the second kind, \( K_\nu(x) \) or \textsc{BesselK}, arises in the separation of variables solutions of partial differential equations~\cite{weisstein2002modified,dunster1990bessel, zhukovsky2017solving}. A key feature of \textsc{BesselK} is its \textit{rapid decay at infinity}. The function diminishes exponentially as \( x \to \infty \), providing a basis for solutions that vanish at large distances. Applications include heat conduction~\cite{hetnarski2009heat}, wave propagation~\cite{gomez2017physics}, and stochastic processes~\cite{pogany2013bessel}.

\textsc{BesselK} also arises in the widely used Mat\'{e}rn kernel, a covariance function for Gaussian processes. The Mat\'{e}rn kernel finds extensive applications across various fields. In spatial statistics, it is used to model spatial correlations~\cite{wang2023parameterization}. In machine learning, it is particularly valuable for Gaussian process regression and Bayesian optimization~\cite{albahar2021robust}. In signal processing, it helps capture correlations in time series and spectral analysis~\cite{yin2020linear}. 
 
At least six different methods for \textsc{BesselK} can be identified in the literature; however, not all of these methods are adopted in existing tools and libraries. (1) \textit{Series Expansions}~\cite{bbtemme}: \( K_\nu(x) \) can be expressed as an infinite series in powers of \( x \), which is truncated to a finite number of terms when \( x \) is small. Most existing libraries utilize series expansions for small \( x \), including MATLAB, Mathematica, GSL, SciPy, and Maple. (2) \textit{Continued Fractions}~\cite{amos1974computation}: \( K_\nu(x) \) can be represented as an infinite fraction with a recursive structure, particularly suitable when \( x \) is moderate or large. This method is also used in most existing libraries for moderate $x$ values. (3) \textit{Asymptotic Expansions}~\cite{olver2009bessel}: This method focuses on the dominant behavior of \( x \) as it becomes large, ignoring lower-order terms, which reduces the accuracy for small \( x \). This method is also used in most existing libraries for large \( x \) values.
(4) \textit{Integral Representations}~\cite{bbtakekawa,luke2014integrals}: \( K_\nu(x) \) is expressed as integrals. This method is flexible and can work with small or large values of $x$. Some libraries that rely on this method include Mathematica, SciPy, and GSL. (5) \textit{Polynomial Fitting}~\cite{grosswald1978bessel}: Polynomial fitting approximates \( K_\nu(x) \), by fitting polynomials to precomputed values of the function over restricted ranges of \( x \) and \( \nu \). (6) \textit{Differential Equation Solvers}~\cite{carley2013numerical}: Numerical methods for solving differential equations, such as finite difference schemes and Runge-Kutta methods~\cite{butcher2007runge}, can be employed to compute \( K_\nu(x) \). These solvers approximate the solution of \( K_\nu(x) \) by discretizing the domain and applying iterative algorithms to solve the governing differential equation.

\section{Background}
This section addresses the challenges in the modeling of Gaussian processes and highlights the importance of accelerating the computations of the \textsc{BesselK} functions to enhance the modeling procedure. Furthermore, we provide an overview of the \emph{ExaGeoStat} software and its implementation, as a testbed to evaluate our novel \textsc{BesselK} implementation.

\subsection{ Gaussian Processes and Mat\'{e}rn Kernel}
 Gaussian process (GP) modeling is a probabilistic machine learning approach that defines distributions over functions and distinguished by its mean function $m(\mathbf{x})$, and covariance function $C(\mathbf{x}, \mathbf{x}')$, where $\mathbf{x}$, $\mathbf{x}'$ are locations. A commonly used method in Gaussian process modeling is the maximum likelihood estimation (MLE). In MLE, an optimization process iterates over a given log-likelihood function to estimate statistical parameters that describe the underlying data. This is achieved by constructing a positive definite covariance matrix that captures the correlation between different observations. The log-likelihood function \( \mathcal{L}( \boldsymbol{\theta}) \) is represented as:
\vspace{-1mm}
\[
\mathcal{L}( \boldsymbol{\theta}) = -\frac{1}{2} \left[ N \log(2\pi) + \log\left(|\mathbf{\Sigma}( \boldsymbol{\theta})|\right) + \mathbf{z}^\top \mathbf{\Sigma}(\boldsymbol{\theta})^{-1} \mathbf{z} \right],
\]
\noindent
 where \( \boldsymbol{\theta} \) is a set of parameters to be fit, \( N \) is the number of observations, \( \mathbf{\Sigma}(\boldsymbol{\theta}) \) is the parameterized covariance matrix, \( |\mathbf{\Sigma}(\boldsymbol{\theta})| \) is its determinant, and \( \mathbf{z} \) is the vector of observed data. The Matérn function serves as a fundamental component in spatial statistics, representing a significant domain where Gaussian processes demonstrate their utility, owing to its adaptability in characterizing spatial correlation structures. Its parametric flexibility enables precise modeling of spatial relationships across diverse applications and datasets~\cite{wang2023parameterization}.

The Mat\'{e}rn function can be represented as:
\[C(\mathbf{x}, \mathbf{x}')={\cal M}(r;\boldsymbol{\theta}) = \frac{\sigma^2}{2^{\nu - 1}\Gamma(\nu)}\left(\frac{r}{\beta}\right)^{\nu}K_{\nu}\left(\frac{r}{\beta}\right),\]
\noindent
where \( r = \| \mathbf{x} - \mathbf{x}' \| \) is the distance between two spatial locations and $\boldsymbol{\theta} = (\sigma^2, \beta, \nu)^\top$, \( \nu > 0 \) is a parameter that controls the smoothness of the function, \( \beta \) is the length scale parameter, \( \sigma^2 \) is the variance, \( \Gamma(\nu) \) is the Gamma function, and \( K_\nu(\cdot) \) is the modified Bessel function of the second kind of order \(\nu\) (\textsc{BesselK($x, \nu$)}). 

The \textsc{BesselK} is the core in the generation of the covariance matrix when using the Mat\'{e}rn function,
which is required not only during the modeling process but also for prediction tasks and the generation of synthetic datasets~\cite{gramacy2020surrogates, leandro2021exploiting}.


\subsection{ExaGeoStat: A Parallel Tile-Based Framework for Geospatial Data Analysis}


\emph{ExaGeoStat} is a high-performance software package for large-scale climate and environmental geostatistics~\cite{abdulah2018exageostat}. It evaluates the log-likelihood function for spatial datasets using a range of covariance models, including the Mat\'{e}rn covariance function. This enables efficient parameter estimation and prediction for large-scale spatial datasets by leveraging state-of-the-art dense linear algebra libraries (e.g., CHAMELEON~\cite{chameleon2024}, and DPLASMA~\cite{bosilca2011flexible}) and runtime systems (e.g., StarPU~\cite{augonnet2009starpu}, and PaRSEC~\cite{hoque2017dynamic}). \emph{ExaGeoStat} achieves high performance across diverse hardware architectures, including multicore CPUs, GPUs, and distributed systems. The software supports both exact and approximate computations~\cite{abdulah2018parallel,abdulah2019geostatistical,cao2023reducing,zhang2024parallel}.

\emph{ExaGeoStat} relies on tile-based algorithms to effectively leverage the underlying runtime system to distribute tasks across available hardware resources. The runtime system efficiently schedules the processing of matrix tiles across computational units, maximizing performance and resource utilization. All executions are conducted exclusively on GPUs to further optimize the performance of \emph{ExaGeoStat}. 


\section {Algorithm for \textsc{BesselK}}
Our \textsc{BesselK} algorithm is built to compute a reasonable range of \( x \), and \( \nu \) that covers a wide spectrum of applications builds upon two existing methods: Temme's series expansion and Takekawa's algorithm, reviewed, respectively, in the following two subsections.


\subsection{Temme's series expansion}

Most existing libraries rely on series expansions to compute the \textsc{BesselK} function when \( x \) is small~\cite{bbtemme}. Series expansions offer an efficient and accurate approximation in this regime, with the series expressed as:
\begin{equation}
    K_\nu(x) = \sum_{k=0}^\infty c_k f_k, \quad K_{\nu+1}(x) = \frac{2}{x} \sum_{k=0}^\infty c_k h_k,
    \label{eq:temme_main}
\end{equation} where
\[
c_k = \frac{(x^2 / 4)^k}{k!}, \quad h_k = -k f_k + p_k,
\]
\noindent
and \( p_k \) and \( q_k \) are recurrence relations defined as:
\[
p_k = \frac{p_{k-1}}{k - \nu}, \quad q_k = \frac{q_{k-1}}{k + \nu}.
\]
\noindent
The term \( f_k \) is computed iteratively:
\[
f_k = \frac{k f_{k-1} + p_{k-1} + q_{k-1}}{k^2 - \nu^2}.
\]
\noindent
The recurrence relations are initialized with:
\begin{equation}
    p_0 = \frac{1}{2} \left(\frac{x}{2}\right)^{-\nu} \Gamma(1 + \nu), \quad q_0 = \frac{1}{2} \left(\frac{x}{2}\right)^\nu \Gamma(1 - \nu),
\label{eq:temme_initpq}
\end{equation}
\begin{equation}
    f_0 = \frac{\nu \pi}{\sin(\nu \pi)} \left[ \cosh(\sigma) \Gamma_1(\nu) + \frac{\sinh(\sigma)}{\sigma} \ln\left(\frac{2}{x}\right) \Gamma_2(\nu) \right],
    \label{eq:temme_initf}
\end{equation}
where \( \sigma \) and \( \Gamma_1(\nu) \), \( \Gamma_2(\nu) \) are precomputed constants for the expansion. 

The original formulation of Temme's series expansion in~\cite{bbtemme} does not provide sufficiently stable or accurate results for \( \nu \geq 1.5 \). The direct evaluation of \( p_0, q_0, \text{ and } f_0 \) for large orders of \( \nu \) increases the error of this approximation method. To address these limitations and improve numerical stability and accuracy, Campbell~\cite{campbell1980temme} proposed leveraging the recurrence relation for large-order \( \nu = \mu + M \) (where \( M = \lfloor \nu + 0.5 \rfloor \)) of \( K_\nu(x) \). 

This approach uses starting values \( K_\mu(x) \) and \( K_{\mu+1}(x) \), with \( -\frac{1}{2} \leq \mu < \frac{1}{2} \), and applies the forward recurrence relation:
\begin{equation}
K_{\eta+1}(x) = \left(\frac{2\eta}{x}\right)K_\eta(x) + K_{\eta-1}(x). \label{eq:recurrence}
\end{equation}

We employ Temme's series expansion and the recurrence relation to evaluate the \textsc{BesselK} function in the small-\(x\) regime (\(x < 0.1\)).

\subsection{Integral-based Algorithm (Takekawa's Approach)}

In~\cite{bbtakekawa}, Takekawa introduced a method to compute the \textsc{BesselK} function using an integral approach. \( K_{\nu}(x) \) is represented by the following integral from~\cite{watson1922treatise}:
\begin{equation}
K_\nu(x) = \int_0^\infty e^{-x \cosh(t)} \cosh(\nu t) \, \mbox{d}t \overset{\Delta}{=} \int_0^{\infty} f_{\nu, x}(t) \, \mbox{d}t,
\label{eq:integral}
\end{equation}
where \( x > 0 \) and \( \nu \in \mathbb{R} \). This representation is computationally intensive but provides accurate results, especially for small and moderate \( x \).

Takekawa worked with the logarithm of the integrand
\(e^{-x \cosh(t)} \cosh(\nu t)  \label{eq:ft}\). Namely,
\begin{equation}
g_{\nu, x}(t) = \log \cosh(\nu t) - x \cosh(t).
\label{eq:logintegrand}
\end{equation}

The first-order, and second-order derivatives of \( g_{\nu, x}(t) \) with respect to \( t \) are given by:
\[
    g'_{v,x}(t) = v \tanh(vt) - x \sinh(t), 
\]
 \[   g''_{v,x}(t) = v^2 \operatorname{sech}^2(vt) - x \cosh(t). \]

At \( t = 0 \), we have \( g_{\nu,x}(0) = -x \) and \( g'_{\nu,x}(0) = 0 \). Consequently, if \( \nu^2 \leq x \), the function \( g_{\nu,x}(t) \) will always decrease, implying that the maximum value of the function in Equation~\eqref{eq:logintegrand} occurs at \( t = 0 \), as \( g''_{\nu,x}(0) \leq 0 \). In contrast, if \( \nu^2 > x \), the function in Equation~\eqref{eq:logintegrand} will reach its maximum at some \( t \geq 0 \), since \( g''_{\nu,x}(0) > 0 \). More details are provided in~\cite{bbtakekawa}.

To integrate Equation~\eqref{eq:integral}, the region where the maximum value of \( t \), denoted as \( t_{\text{max}} \), is located, can be defined as the region where \( {f_{\nu,x}(t) \geq \epsilon_{\text{machine}} f_{\nu,x}(t_{\text{max}})} \), where \(\epsilon_{\text{machine}}\) is the machine epsilon. This region can be expressed as interval \( [t_0, t_1] \), defined as:
\vspace{-2mm}
\begin{equation}
    [t_0, t_1] = \{ t \mid g_{\nu,x}(t) \geq \log(\epsilon_{\text{machine}}) + g_{\nu,x}(t_{\text{max}}) \}.
\end{equation}
Takekawa defines the integral range by determining \( t_{\text{max}} \) from Equation~\eqref{eq:logintegrand}. If \( \nu^2 \leq x \), then \( t_{\text{max}} = 0 \). Otherwise, \( t_{\text{max}} \) can be found by searching within a specific range of the function \( g_{\nu,x}(t) \). Since \( g'_{\nu,x}(t_{\text{max}}) = 0 \) and \( g'_{\nu,x}(t) < 0 \) for $t > t_{\text{max}}$, the range can be defined as \( [2^{m-1}, 2^{m}] \), where \( m \) is the smallest value such that \( g'_{\nu,x}(2^m) < 0 \). For a detailed explanation, refer to the {\it FINDRANGE} algorithm in~\cite{bbtakekawa}. Afterward, the $t_{\text{max}}$ value can be obtained using binary search and Newton methods; refer to the {\it FINDZERO} algorithm in~\cite{bbtakekawa}.

The integration range is also determined using the {\it FINDZERO} algorithm. To compute \( t_0 \) (the lower bound of integration), if \( \nu^2 \leq x \), then \( t_0 = 0 \). Otherwise, the {\it FINDZERO} algorithm is applied to find \( t_0 \) within the range \( [t_0, t_{\text{max}}] \). For \( t_1 \), the {\it FINDZERO} algorithm is used similarly to search within the range \( {[t_{\text{max}} + 2^{m-1}, t_{\text{max}} + 2^m]} \).

After determining \( t_0 \) and \( t_1 \), integration is performed over the range using a fixed number \( b\) of bins. To ensure numerical stability, the \texttt{log\_sum\_exp} function is applied to \( g_{\nu,x}(t_m) \), resulting in:
\begin{equation}
\begin{split}
\log K_{\nu}(x) \approx & \\ g_{\nu,x}(t_{\text{max}}) + & \log \sum_{m=0}^{b} h \exp \Big\{ c_m \big( g_{\nu,x}(t_m) - g_{\nu,x}(t_{\text{max}}) \big) \Big\}.
\end{split}
\label{eq:refined}
\vspace{-30mm}
\end{equation}
Here, the parameters are defined as:
\vspace{-1.5mm}
\[
h = \frac{t_1 - t_0}{b}, \quad t_m = t_0 + mh, 
\]
\vspace{-5mm}
\[
c_0 = c_n = \frac{1}{2}, \quad c_m = 1 \quad (m = 1, \ldots, b-1).
\vspace{-1.5mm}
\]
$K_{\nu}(x)$ can be obtained by taking the exponential.

\subsection{The Proposed Refined Algorithm} 
Existing numerical libraries, such as MATLAB, SciPy, and Boost C++, do not offer GPU support to evaluate the \textsc{BesselK} function, resulting in a significant performance bottleneck for GPU-accelerated scientific applications that depend on this function. Performing \textsc{BesselK} computations on the CPU while executing other tasks on the GPU is highly time-consuming due to the overhead of data transfers between the CPU and GPU and the GPU's superior parallel processing capabilities compared to the CPU. Although Plesner et al.~\cite{bbeth} recently introduced a GPU-accelerated library for \textsc{BesselK} evaluation, its performance gains are limited. Their implementation outperforms GSL only within the parameter range \((x, \nu) \in (150, 4000] \times (150, 4000]\) and exhibits a relatively limited performance outside of this range compared to GSL.  

\begin{figure}[!hbt]
        \centering
        \includegraphics[width=0.33\textwidth]{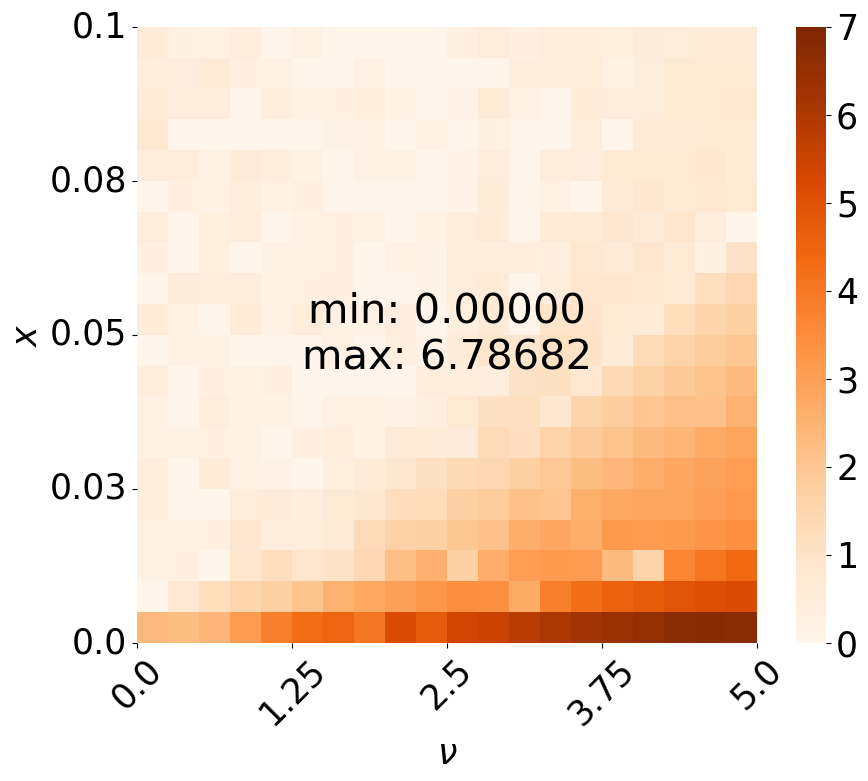} 
        \caption{Relative error of Takekawa's algorithm vs Mathematica for \((\nu, x) \in [0.001, 5] \times [0.001, 0.1]\).}
        \label{fig:Takekawa_heatmap}
    \end{figure}

To address this gap, we propose a novel approach that advances the state-of-the-art in GPU-accelerated \textsc{BesselK} computation, delivering improved performance across a reasonable parameter space. We adopt the quadrature-based algorithm proposed by Takekawa~\cite{bbtakekawa}. Although Takekawa's algorithm demonstrates substantial accuracy for large values of \( x \) and \( \nu \), it notably lacks discussion of cases where \( x < 0.1 \), a range frequently encountered in applications such as spatial statistics. Our analysis reveals that within this range, the integration algorithm exhibits a significant loss of accuracy. Figure~\ref{fig:Takekawa_heatmap} shows a heatmap of the relative error in Takekawa's approach for \( x < 0.1 \), compared to referential results obtained using Mathematica~\cite{paul2018mathematica}. Following Takekawa in ~\cite{bbtakekawa}, we compute the relative error (RE) at each \( x \) and \( \nu \) as:
\[
\text{RE} = 
\log_{10} \left( 1 + \frac{\lvert \text{Mathematica's output} - \text{output} \rvert}
{\varepsilon_{\text{machine}}} \right)
\]
where $\varepsilon_{\text{machine}} = 2^{-52} \approx 2.22 \times 10^{-16}$ for double-precision numbers.

We propose a novel algorithm for the \textsc{BesselK} function that extends the range of \( x \) values beyond those supported by Takekawa's method. We refer to this enhanced approach as the \emph{Refined Algorithm}. Furthermore, we provide an efficient GPU-based implementation for generating large matrices using \textsc{BesselK} within the \emph{ExaGeoStat} software that relies on the \emph{StarPU} runtime system. Our key contributions to improving Takekawa's algorithm are summarized as follows:

1) Our algorithm simplifies the integral in Equation~\eqref{eq:integral} by setting the lower bound to \(0\), as reducing the range by computing \(t_0\) and \(t_1\) is more computationally expensive than extending the range on the GPU. 2) The upper bound is set to a maximum value determined through an empirical bound finder, ensuring applicability across all \(x\) and \(\nu\) ranges. 3) Instead of using the \textit{FINDZERO} algorithm to find a global \(t_{\text{max}}\), as in Takekawa's method, we compute a local \(t_{\text{lmax}}\) for each division within the integral range, enabling faster computation. 4) To improve accuracy, we increase the number of bins \(b\) in the integral, which offsets the expanded range \([t_0, t_1]\) and leverages GPU computational power with minimal performance impact. 5) For \(x < 0.1\), we combine the integral algorithm with Temme's expansion, reducing the relative error from \(6.78682\) in Takekawa's method to \(0.73180\) in the refined algorithm, provided a sufficient number of bins.

\begin{algorithm}
\caption{Empirical Upper Bound Finding}
\begin{algorithmic}[1]
\small
\State \textbf{Given:} $\mathcal{X} \times \mathcal{V} = [0, 140] \times (0, 20]$, $\text{MBK}(x, \nu)$ the reference level of \textsc{logBesselK} using Mathematica, and $\text{RBK}(x, \nu) := \log\left[\int_0^{L} e^{-x \cosh(t)} \cosh(\nu t) \, \mbox{d}t\right]$, where the integration follows the Equation \eqref{eq:refined}.
\ForAll{$(x, \nu) \in \mathcal{X} \times \mathcal{V}$}
\State $\text{AE}(x, \nu) \gets \vert \text{MBK}(x, \nu) - \text{RBK}(x, \nu)\vert$ (Absolute Error for $(x, \nu)$)
\EndFor
\State \(t_1 \gets \min L \;\; s.t. \; \max_{x,\nu} \text{AE}(x, \nu) \leq 10^{-9}\)
\end{algorithmic}
\label{alg:emprical}
\end{algorithm}
For $x \in [0.1, \infty)$, we use a refined version of the integration algorithm proposed in \cite{bbtakekawa} to evaluate \textsc{BesselK}. Instead of determining the integration range $[t_0, t_1]$ dynamically based on $x$ and $\nu$, we fix $t_0 = 0$ and $t_1 = 9$. We set $t_0 = 0$ to match the starting point of the integral, and through the empirical upper bound finding algorithm, we establish $t_1 = 9$ as the optimal endpoint for parameters within $(x, \nu) \in [0, 140] \times (0, 20]$, which corresponds to the expected parameter range in geospatial applications involving the Mat\'{e}rn kernel. To elaborate on, in geospatial and machine learning studies, we can always rescale the 2D plane inside a unit square, whose maximum distance is \(\sqrt{2}\). The typical starting point of optimizing the \(\beta\) parameter in the Mat\'{e}rn kernel is \(0.01\), which gives us the maximum \(x\) we have \(\sqrt{2}/0.01 \approx 140\). We are choosing \(x \in [0, 140]\). The smoothness parameter \(\nu\) controls the smoothness of the data and has mathematical interpretation as the \(m\)-th differentiability for integers $m<\nu$. The case \(\nu > 20\) can be approximated by the squared exponential kernel. Thus, choosing this study region is reasonable. Empirically finding the upper bound is shown in Algorithm~\ref{alg:emprical}.
This broader integration range may slightly increase computational cost but improves GPU efficiency by avoiding performance-degrading conditional branching. Besides, this unified upper bound, no matter what \((x, \nu)\), largely decreased the computational burden compared to the original method, which determines the lower and upper bound for each \((x, \nu)\) pair using Newton-based zero-finders.

Algorithm~\ref{alg:refined} provides a detailed explanation of the steps of our proposed algorithm. The inputs are the pair ($x$, $\nu)$ to compute $K_{\nu}(x)$. To define $b$, the number of bins, increasing it can improve accuracy, but at the cost of reduced performance, as the algorithm must identify the local maximum point $t_{\text{lmax}}$ for each division. However, we observed that fixing the number of bins to $40$ provides a balance, achieving an accuracy threshold that ensures computational stability across different values of $x$ and $\nu$. In Algorithm~\ref{alg:refined}, Temme's expansion method is used to compute the Bessel function if $x\leq 0.1$ (lines 3-7); otherwise, use Equation (\ref{eq:refined}) to compute the Bessel function using $b$ bins (lines 8-13).
\begin{algorithm}
\caption{Refined Algorithm (\textsc{BesselK}($x$, $\nu$))} 
\begin{algorithmic}[1]
\small
\State \textbf{Input:} \((x, \nu) \in \mathcal{X} \times \mathcal{V}\), where $\mathcal{X} \times \mathcal{V}$ is the problem region of evaluation of \textsc{BesselK}
\State \textbf{Given:} $b$ number of bins for numerical integration
\BeginBox[draw=black,dashed]
\If{\(0 \leq x < 0.1\) \textbf{and} \(\nu \in \mathcal{V}\)}
    \State Set $M = \lfloor \nu + 0.5 \rfloor$ and use $\mu = \nu - M$
    \State Use Equation \eqref{eq:temme_initpq} and Equation \eqref{eq:temme_initf} to initialize $p_0, q_0, f_0$ for smoothness $\mu$
    \State Set $15000$ instead of $\infty$ for the sum of Temme's series for $K_{\mu}(x)$ in Equation \eqref{eq:temme_main}
    \State Use the recurrence relation 
    $$K_{\eta+1}(x) = \left(\frac{2\eta}{x}\right)K_\eta(x) + K_{\eta-1}(x)$$ until $\eta + 1 = \nu$ to acquire $K_{\nu}(x)$
\EndBox
\BeginBox[draw=red,dashed]
\Else{ \(x \in \mathcal{X} \backslash [0, 0.1)\) \textbf{and} \(\nu \in \mathcal{V}\)}
    \State $t_0 \gets \text{LB}$ (fixed to $0$)
    \State $t_1 \gets \text{UB}$ (empirical upper bound for all $(x, \nu)$)
    \State $t_{\text{lmax}} \gets \max_{i = 0, \ldots, b}t_i$
    \State Use \begin{align*}
       & \log K_{\nu}(x) \gets  \ g_{\nu,x}(t_{\text{lmax}}) + \\ 
& \log \sum_{m=0}^{b} h \exp \Big\{ c_m \big( g_{\nu,x}(t_m) - g_{\nu,x}(t_{\text{lmax}}) \big) \Big\},
    \end{align*} where $h, t_m, c_m$ are given in Equation~\eqref{eq:refined} and $g_{\nu,x}(\cdot)$ given in Equation \eqref{eq:logintegrand}
    \State $K_{\nu}(x) \gets \exp(\log(K_{\nu}(x)))$
\EndBox
\EndIf
\State \textbf{Output:} \(\textsc{BesselK}(x, \nu) \gets K_{\nu}(x)\)
\end{algorithmic}
\label{alg:refined}
\end{algorithm}

Algorithm~\ref{alg:tile_matern} presents the pseudocode for the CUDA algorithm used to generate a single tile of the covariance matrix based on the Matérn covariance function and the \textsc{BesselK} function, using the refined algorithm. Using the \emph{StarPU} runtime system, the matrix is partitioned into smaller tiles and each tile is assigned to a different GPU to dynamically compute the full covariance matrix, while handling the data transfer from host-to-device and device-to-host. Memory allocation is handled via \emph{StarPU} to optimize data transfer performance between the CPU (host) and the GPU. In lines 5-13, each GPU thread processes a single tile value to compute the \textsc{BesselK} function based on the corresponding \(x\) value, as described in Algorithm~\ref{alg:refined}.

\begin{algorithm}
\small
\caption{GPU Single-Tile Mat\'{e}rn Covariance Generation Algorithm}
\begin{algorithmic}[1]
\Function{GenerateMat\'{e}rnCovariance}{$\mathbf{\ell}^1_x$, $\mathbf{\ell}^1_y$, $\mathbf{\ell}^2_x$, $\mathbf{\ell}_y^2$, $\sigma$, $\beta$, $\nu, m, n$} where $\ell^1$  and $\ell^2$ represents location vectors.
    \State Initialize CUDA grid with dimension $(m + 8- 1 // m) * (n + 8 - 1 // n)$ and block dimensions $8 \times 8$ (which means $64$ threads for each block)
    \State Use \texttt{starpu\_malloc()} for efficient CUDA memory allocation, enabling fast CPU-GPU and GPU-GPU transfers.
    \State Copy location vectors and parameters to GPU memory
    \For{each thread $(i,j)$ in parallel}
        \If{$i < m$ and $j < n$}
            \State $c \gets \sigma^2 / (2^{\nu-1}\Gamma(\nu))$
            \State $d \gets \sqrt{(\mathbf{\ell}^2_x[j] - \mathbf{\ell}^1_x[i])^2 + (\mathbf{\ell}^2_y[j] - \mathbf{\ell}^1_y[i])^2)}$
            \State $r \gets d/\beta$
            \If{$r = 0$}
                \State $A[i + j\times m] \gets \sigma^2$
            \Else
                \State $A[i + j\times m] \gets c \cdot r^{\nu} \cdot \textsc{BesselK}(r, \nu)$
            \EndIf
        \EndIf
    \EndFor
    \State Transfer back the generated submatrices and parameters to CPU
    \State Synchronize CUDA stream
\EndFunction
\State Free the CUDA memory and destroy the CUDA stream
\end{algorithmic}
\label{alg:tile_matern}
\end{algorithm}

\subsection{GPU Optimizations on Algorithm \ref{alg:refined}}
Algorithm \ref{alg:refined} is written for CUDA where threads are collected in blocks of threads; 64 threads in our case gained good performance. The threads in a block are collected in warps of 32 threads, where each warp is used for the Single Instruction Multiple Threads (SIMT) execution model~\cite{bbeth}. When naively parallelized on GPUs, the \(i\)-th thread computes \(K_{\nu}(x)\) for the \(i\)-th input value \((x_i, \nu_i)\).

Specifically, the covariance matrix we target involves computing the modified Bessel function for \(r/\beta\), where \(r\) represents the distance between two spatial locations, and \(\beta\) is a fixed parameter during a single generation operation. In our implementation, we ordered the spatial locations using Morton's ordering~\cite{ordering2018} to ensure that distances within the same block remain close and consistently follow a single branch of the if-else condition in the CUDA implementation since we are operating in a tile-based manner.

\section{Experimental Results}
In this section, we evaluate the implementation of the refined \textsc{BesselK} algorithm on the GPU within the \emph{ExaGeoStat} software, focusing on its impact in accelerating the computation of individual matrix elements when millions to trillions of elements are computed. The experiments focus on four main objectives: (1) Assessing the accuracy of computing \textsc{BesselK} for specific $x$ and $\nu$ values; (2) Analyzing the overall accuracy of spatial statistical modeling using a synthetic dataset, emphasizing our GPU-based implementation for matrix generation across different spatial correlation levels; (3) Validating the accuracy of the proposed implementation in modeling real datasets within the context of climate and weather applications; (4) Evaluating the performance of full covariance matrix generation within \emph{ExaGeoStat}, using both single and multiple GPUs.

The implementation associated with this research, which functions independently from the \emph{ExaGeoStat} framework and focuses exclusively on optimizing \textsc{BesselK} for the CUDA architecture, is available for public access at: \url{https://github.com/stsds/CuBesselK}

\subsection{Experimental Configuration}
The results for the overall accuracy of the spatial statistical modeling (Figures \ref{fig:mle-combined} -- \ref{fig:iternbins}) are generated using NVIDIA GV100 32GB with Intel Cascade Lake 256GB CPU, each with 100 repetitions. The real dataset application follows the same settings. Performance assessments (Figures \ref{fig:v100comparison} -- \ref{fig:scaling}) use five repetitions for benchmarking. We use NVIDIA GV100 32GB and NVIDIA A100 80GB GPUs with Intel Cascade Lake CPU. For \emph{ExaGeoStat}, we use CUDA 11.8, gcc 11.2.1, CMake 3.24.2, OpenMPI 4.1.4, OneAPI 2022.3, and StarPU 1.3.11.

\subsection{Relative Error Analysis Against Mathematica}

For accuracy assessment, existing work on implementing the \textsc{BesselK} function often uses Mathematica's \texttt{BesselK[nu, x]} as the benchmark. In this study, we compare the performance of the GSL library, Takekawa's algorithm, and the refined algorithm against Mathematica. The relative errors for various values of \( x \) and \( \nu \) are presented as a heatmap. The heatmap spans the region \( (\nu, x) \in [0.001, 20] \times [0.001, 140] \), which adequately covers the parameter range relevant to Gaussian processes.

\begin{figure}[!hbt]
    \centering
    \begin{subfigure}[b]{0.24\textwidth}
        \centering
        \includegraphics[width=\textwidth]{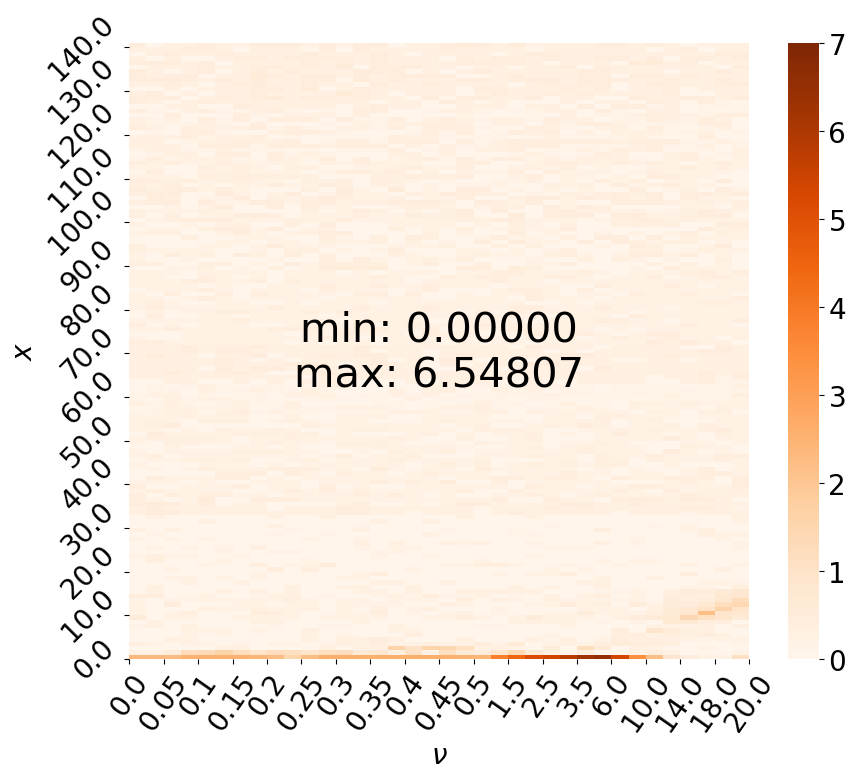} 
        \caption{Takekawa's algorithm.}
        \label{fig:figure1}
    \end{subfigure}
    \hfill
    \begin{subfigure}[b]{0.24\textwidth}
        \centering
        \includegraphics[width=\textwidth]{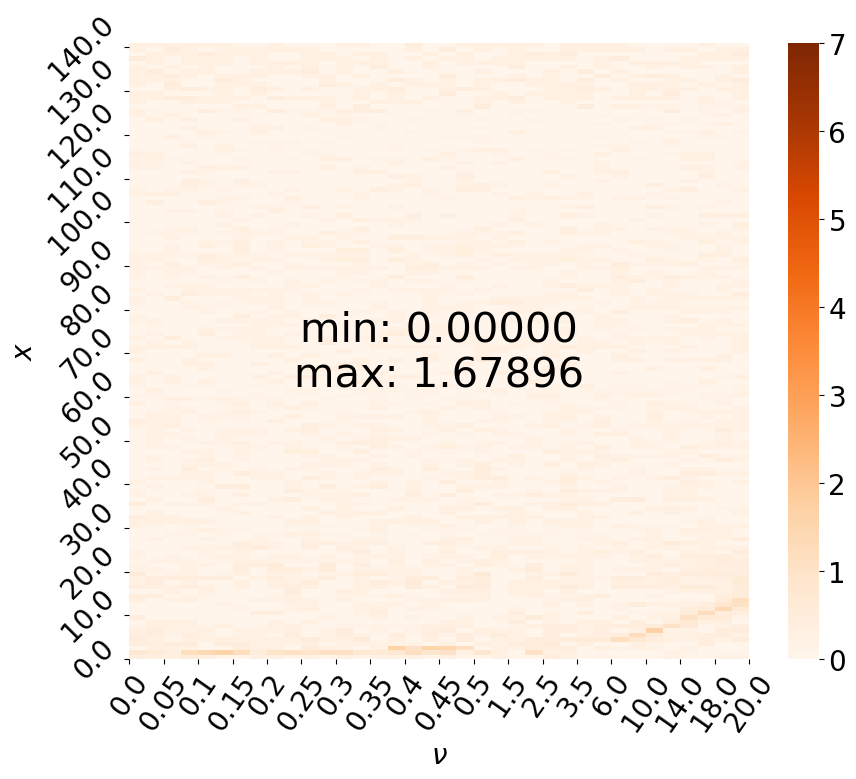} 
        \caption{Refined algorithm.}
        \label{fig:figure2}
    \end{subfigure}
    \hfill
    \begin{subfigure}[b]{0.24\textwidth}
        \centering
        \includegraphics[width=\textwidth]{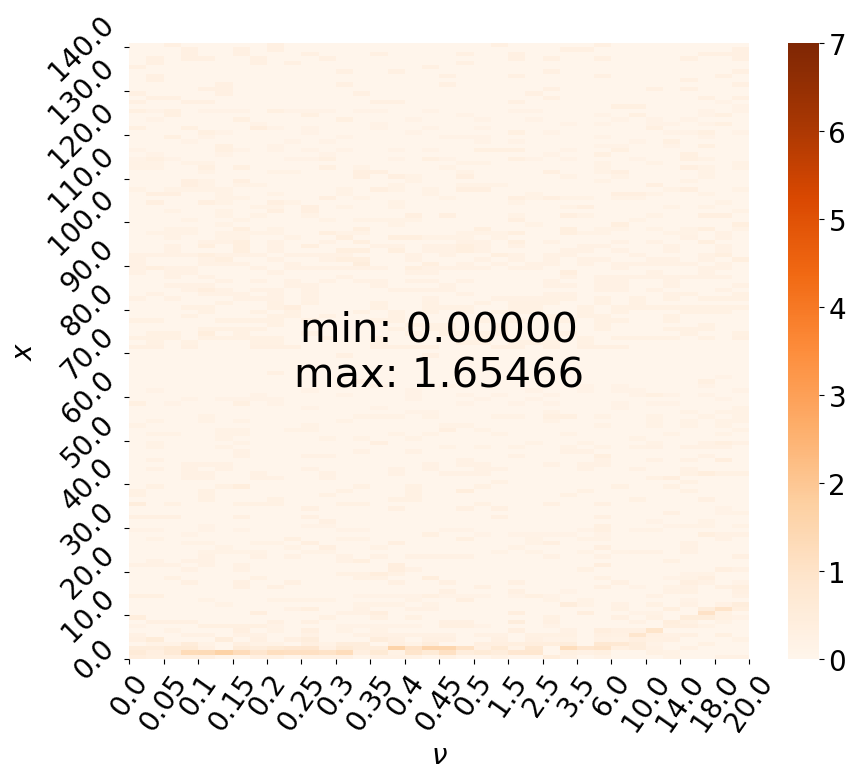} 
        \caption{GNU Scientific Library.}
        \label{fig:figure3}
    \end{subfigure}
    \caption{\textsc{LogBesselK} accuracy comparisons using heatmap for $(\nu, x) \in [0.001, 20] \times [0.001, 140]$.}
    \label{fig:heatmap1}
\end{figure}

\begin{figure}[!hbt]
    \centering
    \begin{subfigure}[b]{0.24\textwidth}
        \centering
        \includegraphics[width=\textwidth]{figs/logk_prec_takekawanew.png} 
        \caption{Takekawa's algorithm.}
        \label{fig:figure5}
    \end{subfigure}
    \hfill
    \begin{subfigure}[b]{0.24\textwidth}
        \centering
        \includegraphics[width=\textwidth]{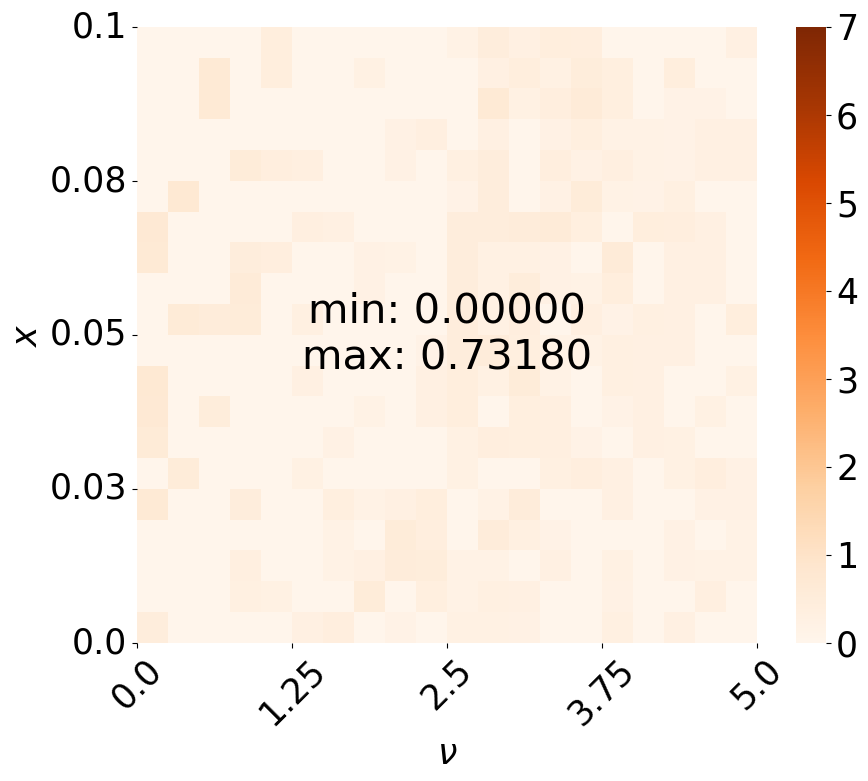} 
        \caption{Refined algorithm. }
        \label{fig:figure6}
    \end{subfigure}
    \hfill
        \begin{subfigure}[b]{0.24\textwidth}
        \centering
        \includegraphics[width=\textwidth]{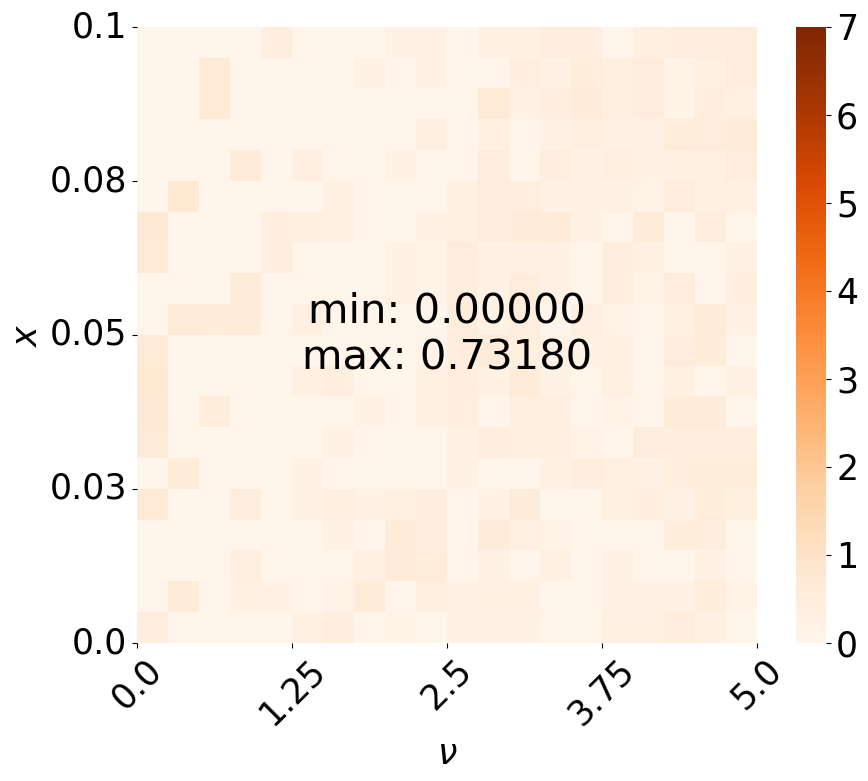} 
        \caption{GNU Scientific Library.}
        \label{fig:figure7}
    \end{subfigure}
    \hfill
    \caption{\textsc{LogBesselK} accuracy comparisons using heatmap for $(\nu, x) \in [0.001, 5] \times [0.001, 0.1]$.}
    \label{fig:heatmap2}
\end{figure}

Figure~\ref{fig:heatmap1} presents three heatmaps for the target region using the three implementations.  As shown, the relative errors of the \textsc{LogBesselK} in Takekawa's algorithm are larger than those of GSL and the refined algorithm, with a maximum error of \( 6.54807 \). This is mainly due to the uncovered region where \( x < 0.1 \). GSL and the refined algorithm exhibit very close relative errors, with values of \( 1.67896 \) and \( 1.65466 \), respectively. 
To highlight the advantages of our method over Takekawa's method for \( x \leq 0.1 \), we further zoom into the region \( (\nu, x) \in [0.001, 5] \times [0.001, 0.1] \) in Figure~\ref{fig:heatmap2}. 

\subsection{Spatial Data Modeling Accuracy within ExaGeoStat}


 Estimating the relative error of the Bessel function for each pair $(x, \nu)$ is crucial to evaluate the effectiveness of a given implementation. However, beyond this, the function \textsc{BesselK}, which is buried within a kernel to generate a matrix, can affect the accuracy of subsequent matrix operations if the accumulated error from different calculations becomes significant. We integrate our implementation into \emph{ExaGeoStat} to generate the Mat\'{e}rn kernel covariance matrix to assess this point. We use the modeling process in \emph{ExaGeoStat}, specifically MLE with gradient-free optimization, which requires multiple iterations to converge and generate a set of estimates of the parameters $\sigma^2$, $\beta$ and $\nu$ that effectively describe the underlying field.


The accuracy of the MLE operation is typically evaluated using simulations in which synthetic data are generated~\cite{salvana2022parallel}. 
We use synthetic data generated within a 2D space, as described in~\cite{sun2016statistically}. We assess the accuracy of the modeling results across three representative scenarios commonly encountered in spatial statistics. These scenarios were characterized by different levels of spatial correlation: weak ($\beta = 0.03$ and $\beta = 0.025$), medium ($\beta = 0.1$ and $\beta = 0.075$) and strong ($\beta = 0.3$ and $\beta = 0.2$), for two different levels of smoothness, rough field ($\nu = 0.5$) and smooth field ($\nu = 1$), respectively. The variance was fixed at $\sigma^2 = 1$ in all experiments.



\begin{figure*}[ht]
    \centering
    \begin{minipage}[t]{0.48\textwidth}
        \begin{subfigure}[b]{\textwidth}
            \centering
            \includegraphics[width=\textwidth]{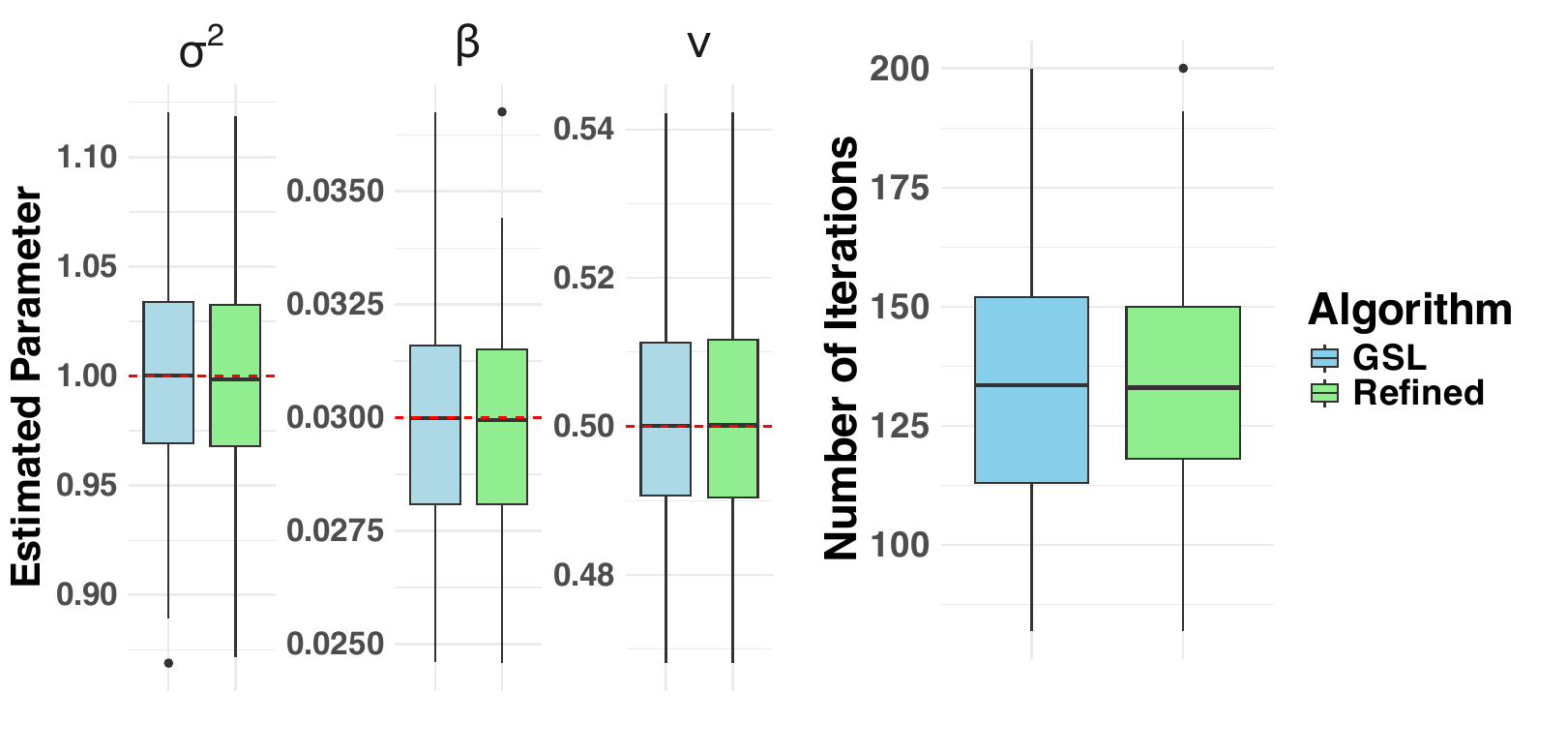}
            \caption{Weak correlation $\beta=0.03$, when $\nu=0.5$.}
            \label{fig:weak0.5}
        \end{subfigure}
        \vfill       
        \begin{subfigure}[b]{\textwidth}
            \centering
            \includegraphics[width=\textwidth]{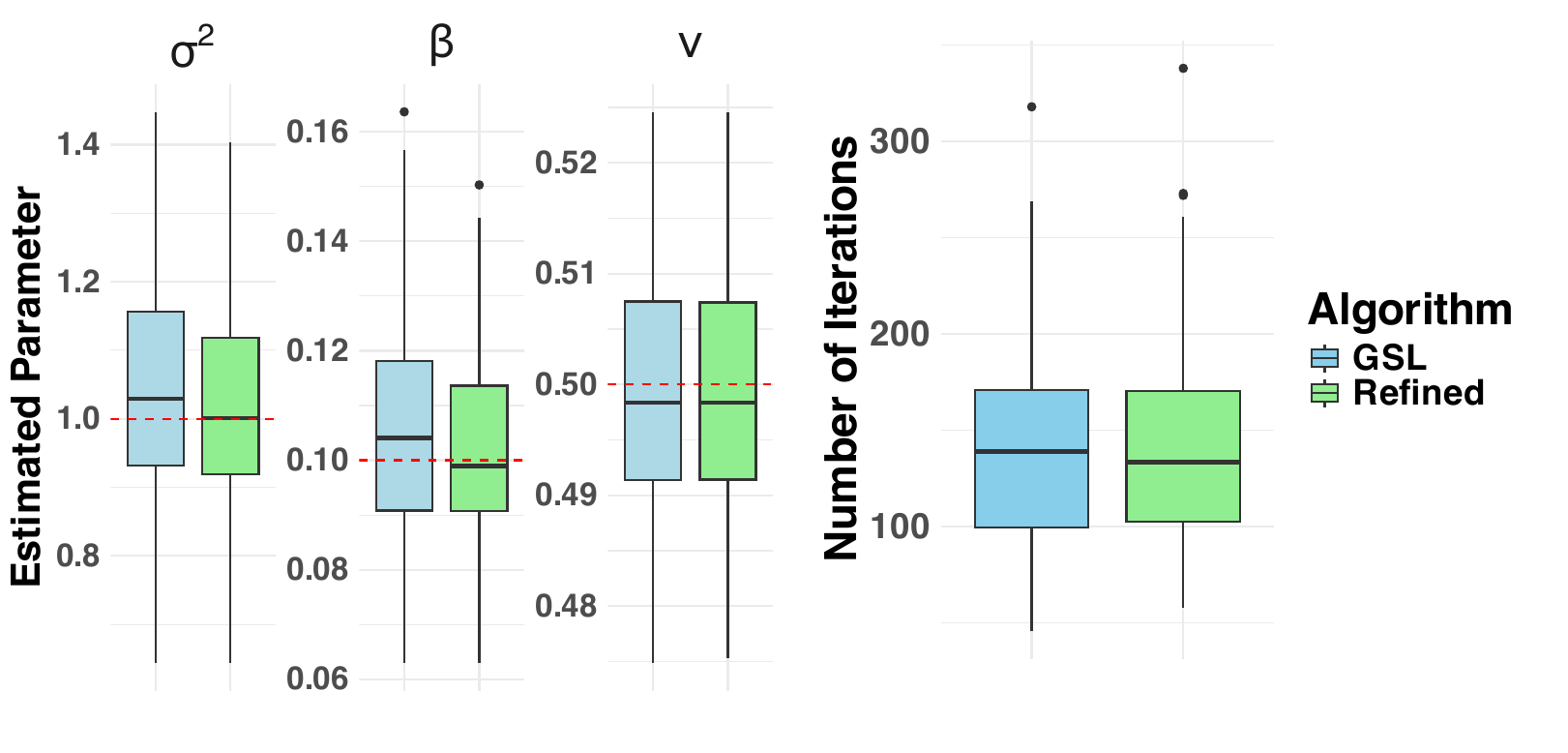}
            \caption{Medium correlation $\beta=0.1$, when $\nu=0.5$.}
            \label{fig:medium0.5}
        \end{subfigure}
 \vfill 
        \begin{subfigure}[b]{\textwidth}
            \centering
            \includegraphics[width=\textwidth]{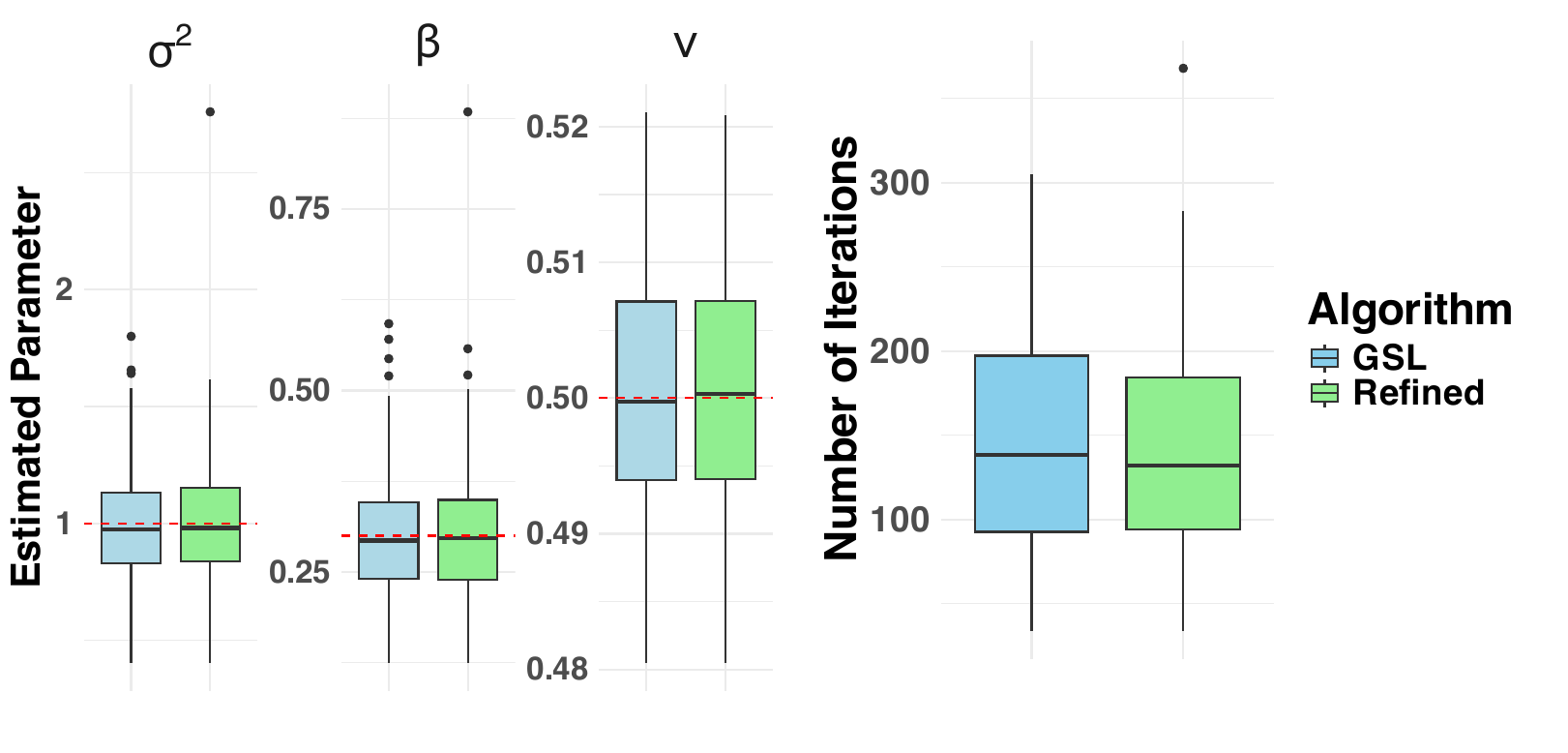}
            \caption{Strong correlation $\beta=0.3$, when $\nu=0.5$.}
            \label{fig:strong0.5}
        \end{subfigure}
    \end{minipage}
    \hfill
    \begin{minipage}[t]{0.48\textwidth}
        \begin{subfigure}[b]{\textwidth}
            \centering
            \includegraphics[width=\textwidth]{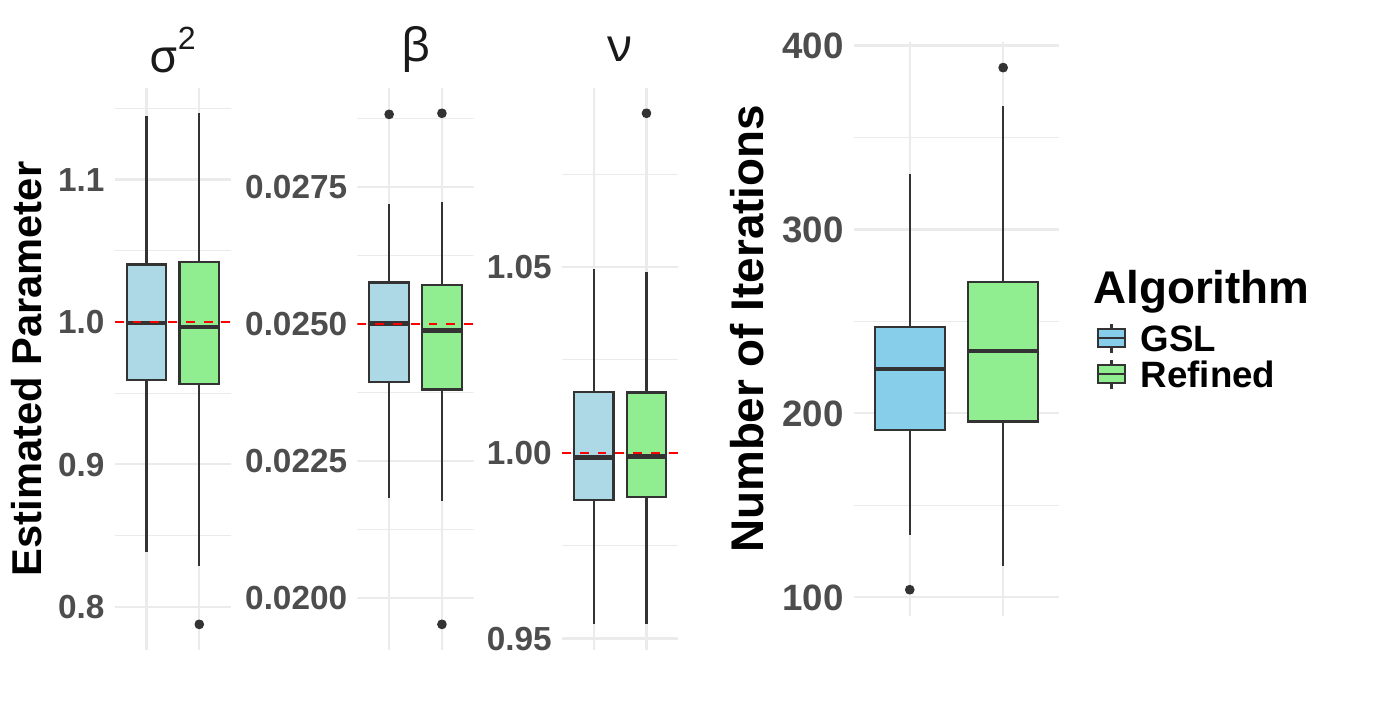}
            \caption{Weak correlation $\beta=0.025$, when $\nu=1$.}
            \label{fig:weak1}
        \end{subfigure}
 \vfill 
        \begin{subfigure}[b]{0.93\textwidth}
            \centering
            \includegraphics[width=\textwidth]{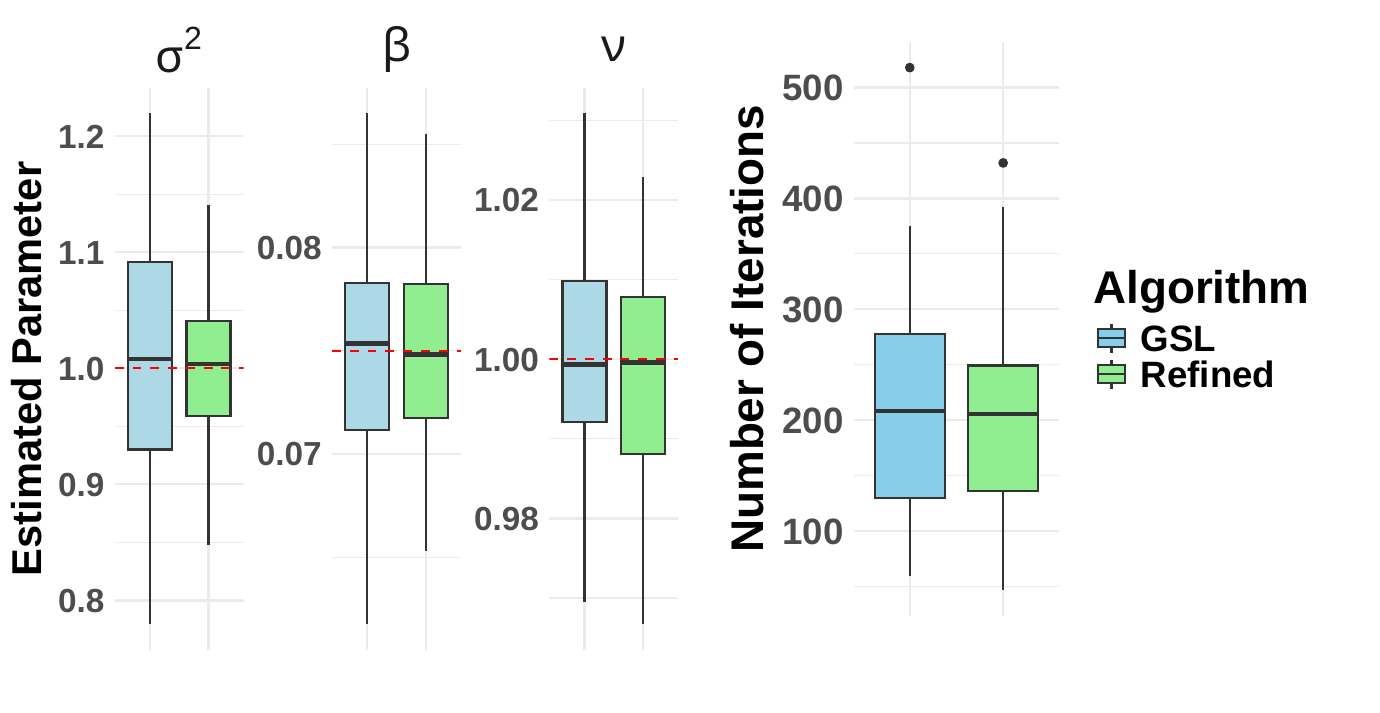}
            \caption{Medium correlation $\beta=0.075$, when $\nu=1$.}
            \label{fig:medium1}
        \end{subfigure}
 \vfill 
        \begin{subfigure}[b]{0.985\textwidth}
            \centering
            \includegraphics[width=\textwidth]{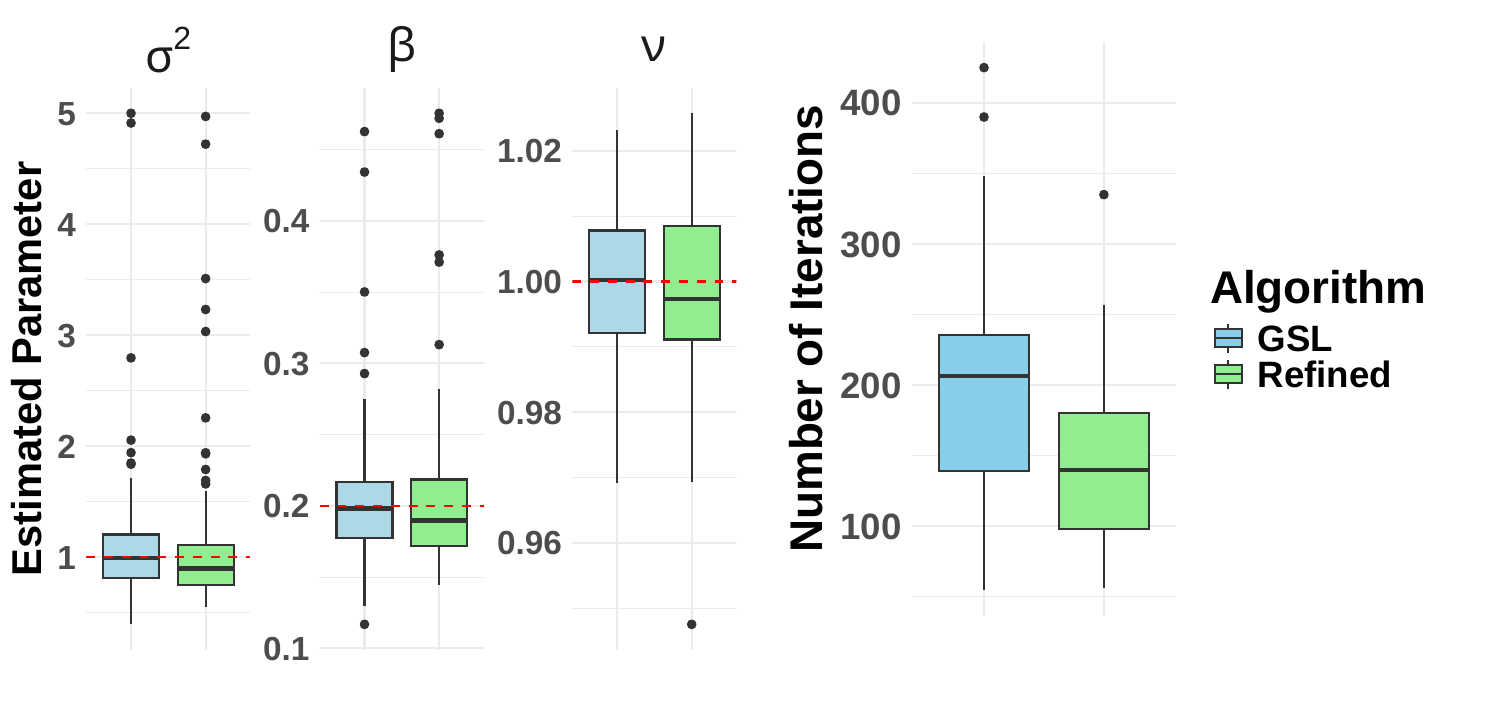}
            \caption{Strong correlation $\beta=0.2$, when $\nu=1$.}
            \label{fig:strong1-16}
        \end{subfigure}
    \end{minipage}
    \caption{Boxplots of MLE optimization results over 100 replicas comparing GSL (CPU) and refined algorithm (GPU). Left column: $\nu=0.5$ cases. Right column: $\nu=1$ cases. All plots show parameter estimates ($\sigma^2$, $\beta$, $\nu$) and iteration counts with red dashed lines indicating true values of parameters.}
    \label{fig:mle-combined}
\end{figure*}

Figures \ref{fig:weak0.5} -- \ref{fig:strong1-16} provide a comparison of parameter estimation using MLE, taking advantage of both GSL and the proposed refined algorithm. The analysis, conducted across three correlation levels (weak, medium, and strong), includes boxplots that illustrate the estimation of three key parameters (\(\sigma^2\), \(\beta\), \(\nu\)) and iteration counts.

In the weak correlation scenario (Figure \ref{fig:weak0.5}), both implementations achieve comparable accuracy in parameter estimation. Furthermore, the average iteration count for the refined algorithm is similar to that of GSL, indicating a comparable computational efficiency. The medium correlation scenario (Figure \ref{fig:medium0.5}) demonstrates increased variability in parameter estimation. Both implementations maintain comparable accuracy, but the refined algorithm exhibits slightly more consistent estimates across all parameters. For strong spatial correlation (Figure \ref{fig:strong0.5}), the most challenging scenario, both implementations exhibit wider parameter distributions, reflecting the increased difficulty of the estimation process. The refined algorithm achieves comparable accuracy and number of iterations.

\begin{figure}[!hbt]
    \centering
    \includegraphics[width=0.5\textwidth]{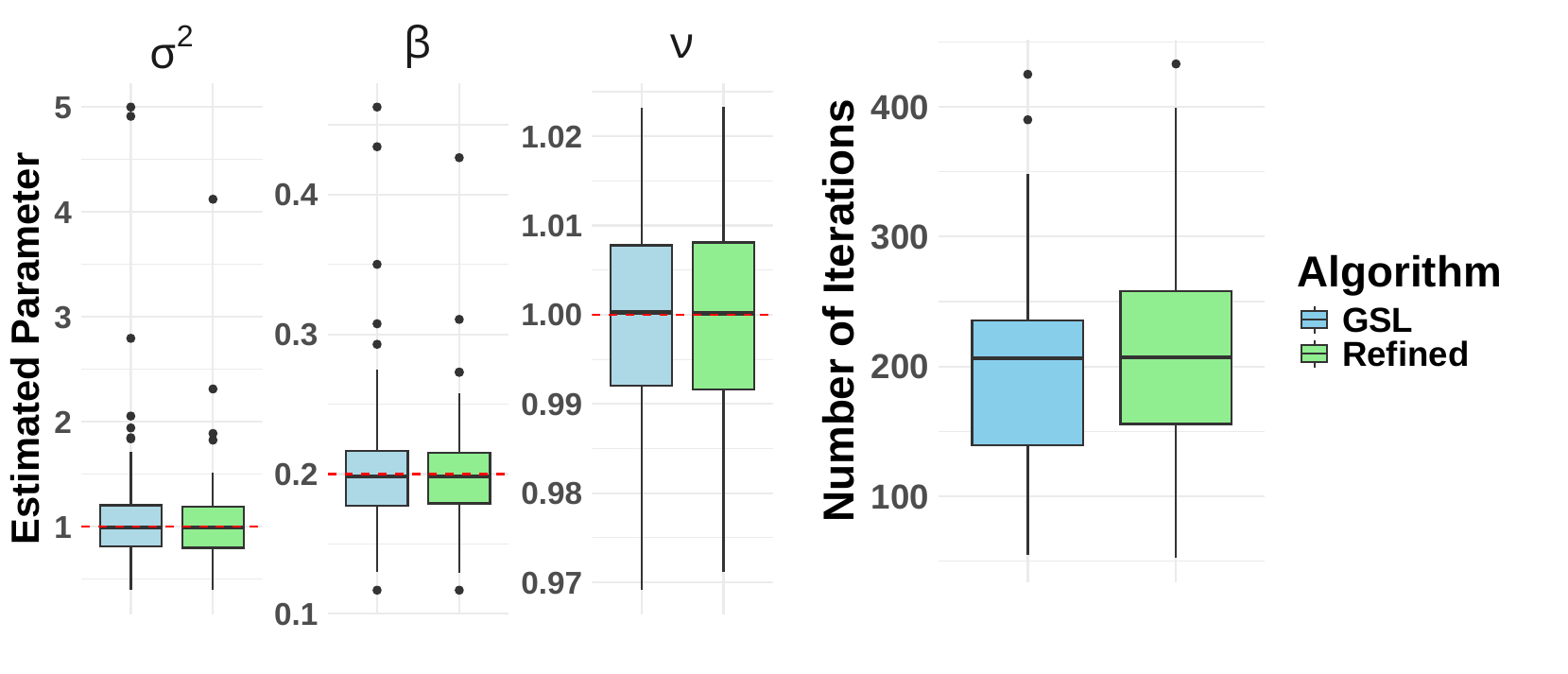}
    \caption{Boxplots of MLE optimization results over 100 replicas using GSL on CPU and the refined algorithm on GPU when  $\nu=1$ and using $b=40$ bins.}
    \label{fig:strong1-40}
\end{figure}

In the smoother case where $\nu = 1.0$, Figures \ref{fig:weak1}, \ref{fig:medium1}, and \ref{fig:strong1-16} demonstrate that the refined algorithm achieves an accuracy similar to GSL. However, it shows a slightly higher bias between the median and ground truth in the strong correlation scenario, which remains within an acceptable range. This is expected, as a larger $\nu$ makes the \textsc{BesselK} approximation more sensitive to the number of bins used for numerical integration. However, the refined algorithm requires significantly fewer iterations than GSL. We also experimented with a strong correlation scenario using $b = 40$ instead of $16$. The results are shown in Figure \ref{fig:strong1-40}. The refined algorithm achieves a level of accuracy similar to that of GSL, and the bias in parameter estimation is eliminated.
These results indicate that the refined algorithm largely preserves accuracy compared to GSL while achieving improved computational efficiency. Despite requiring a similar number of optimization iterations, each iteration is significantly faster, as demonstrated in the next section. Furthermore, consistent performance across varying correlation strengths highlights the robustness of the refined algorithm for spatial statistics and Gaussian process applications. By comparing Figure \ref{fig:strong1-16} and Figure \ref{fig:strong1-40}, it can be concluded that adjusting the number of bins inherently involves a trade-off between accuracy and computational efficiency. Users are advised to carefully consider this balance, particularly in cases where $\nu$ is larger and the spatial correlation is strong.

\subsection{Accuracy Across Different Numbers of Bins}
\begin{figure}[!hbt]
    \centering
    \begin{subfigure}[b]{0.5\textwidth}
        \centering
        \includegraphics[width=\textwidth]{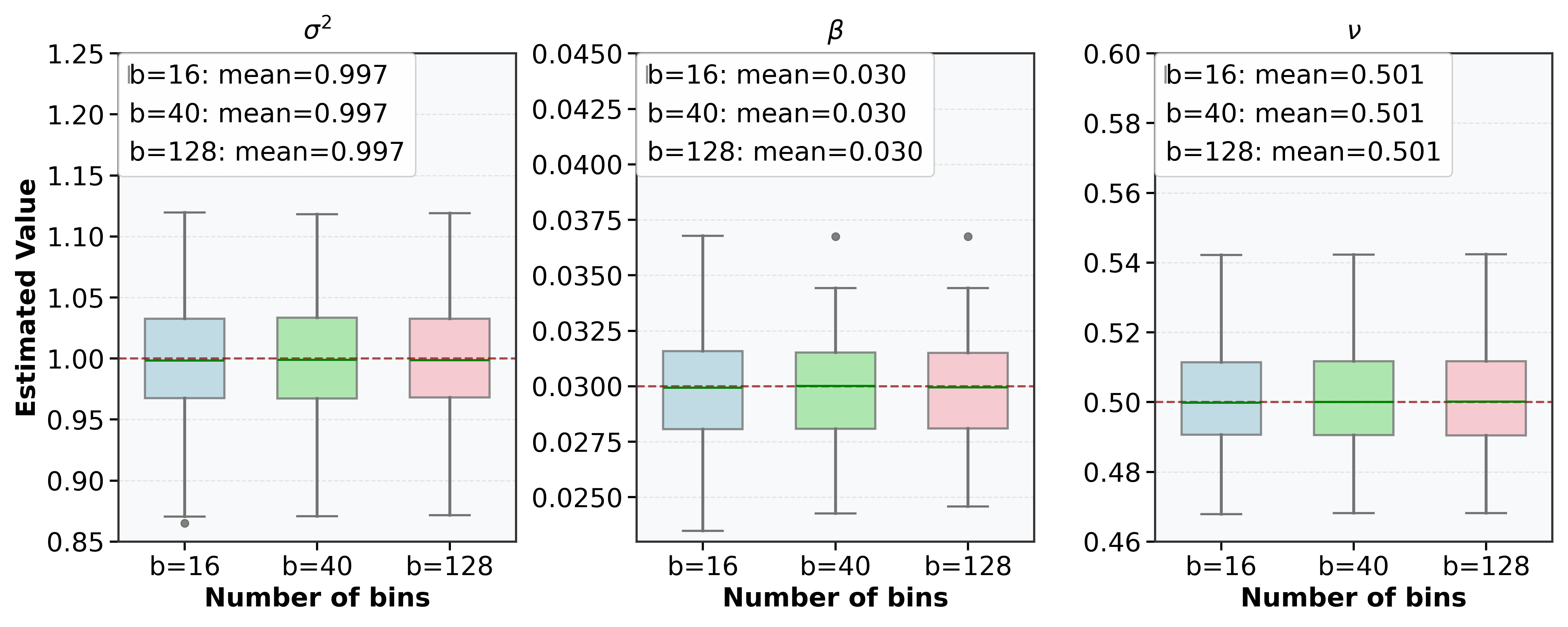} 
        \caption{Weak correlation $\beta=0.03$.}
        \label{fig:mleweak0.5}
    \end{subfigure}
    \vfill
    \begin{subfigure}[b]{0.5\textwidth}
        \centering
        \includegraphics[width=\textwidth]{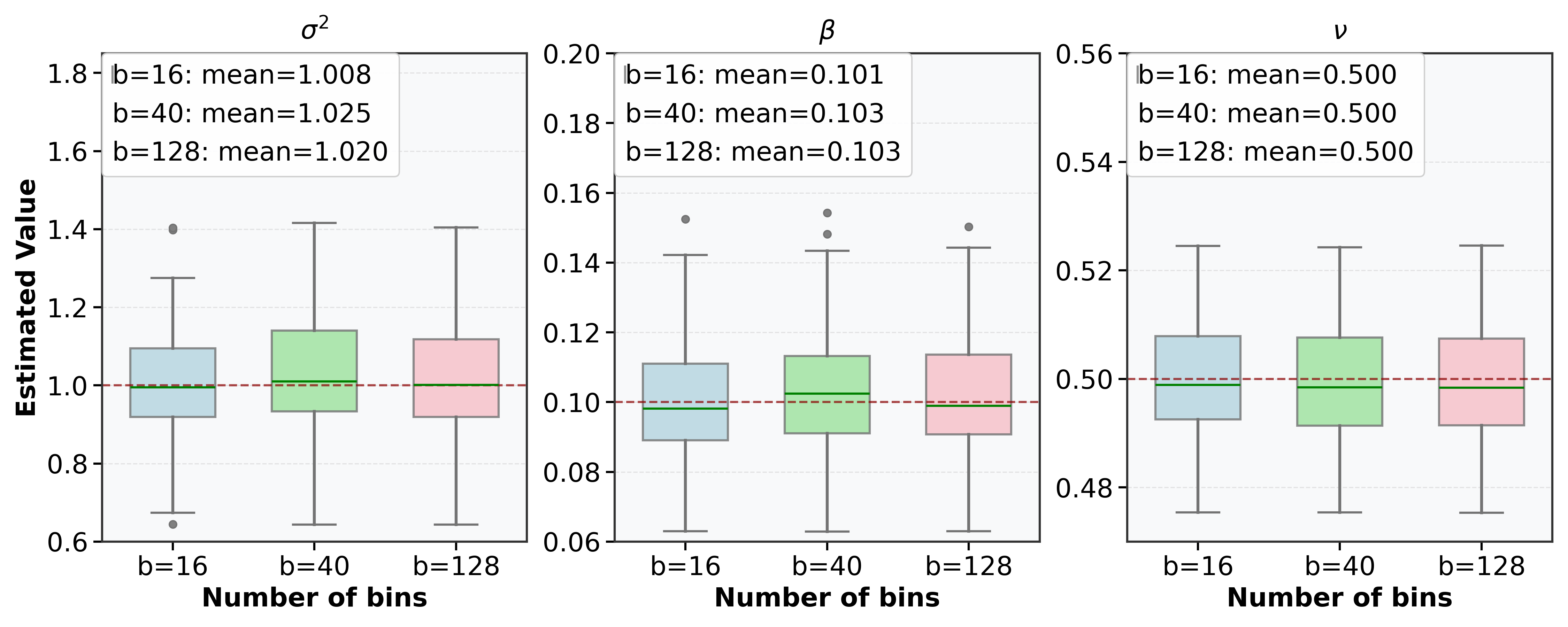} 
        \caption{Medium correlation $\beta=0.1$.}
        \label{fig:mlemedium0.5}
    \end{subfigure}
    \vfill
    \begin{subfigure}[b]{0.5\textwidth}
        \centering
        \includegraphics[width=\textwidth]{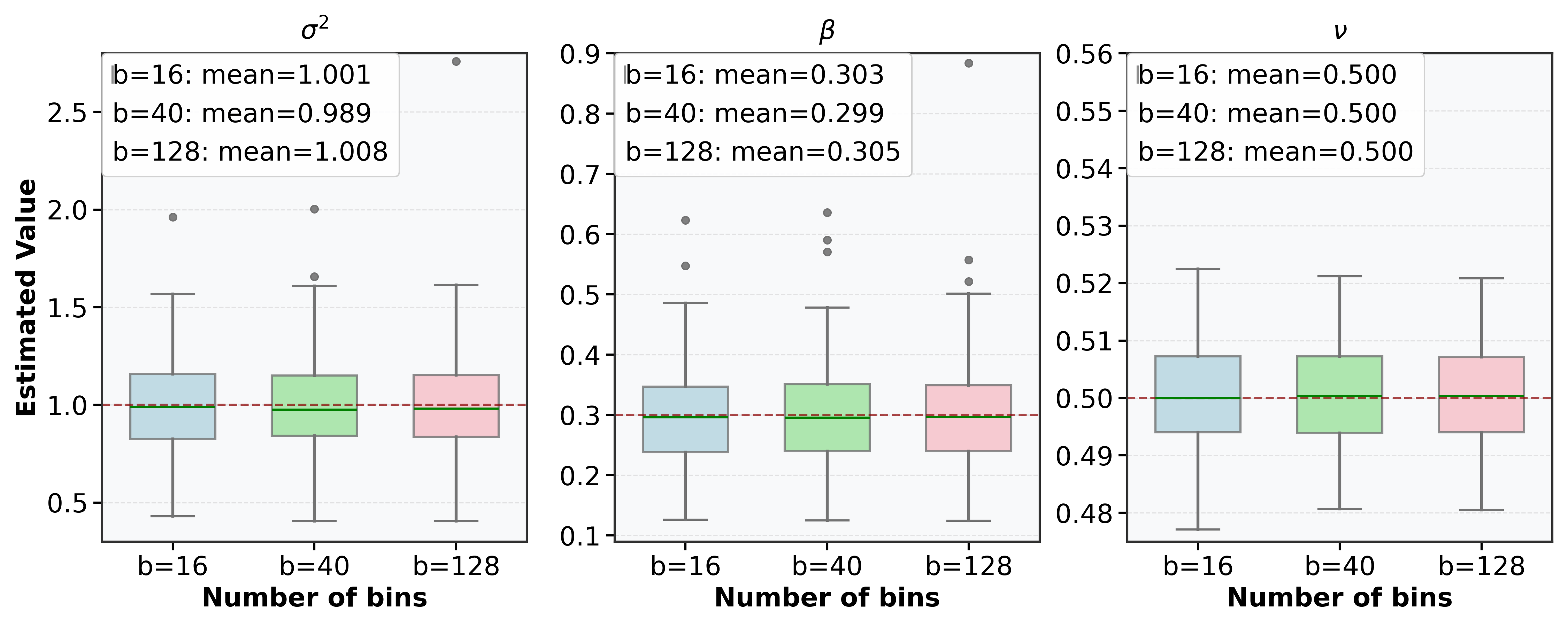} 
        \caption{Strong correlation $\beta=0.3$.}
        \label{fig:mlestrong0.5}
    \end{subfigure}
    \caption{Parameter estimation over 100 replicas using the refined algorithm was evaluated for varying bin counts to estimate \(\sigma^2\), \(\beta\), and \(\nu\) across different correlation levels with problem size $51{,}076$. Red dashed lines: true values of parameters.}
    \label{fig:mlenbins}
\end{figure}
\begin{figure}[!htb]
    \centering
    \begin{subfigure}[b]{0.24\textwidth}
        \centering
        \includegraphics[width=\textwidth]{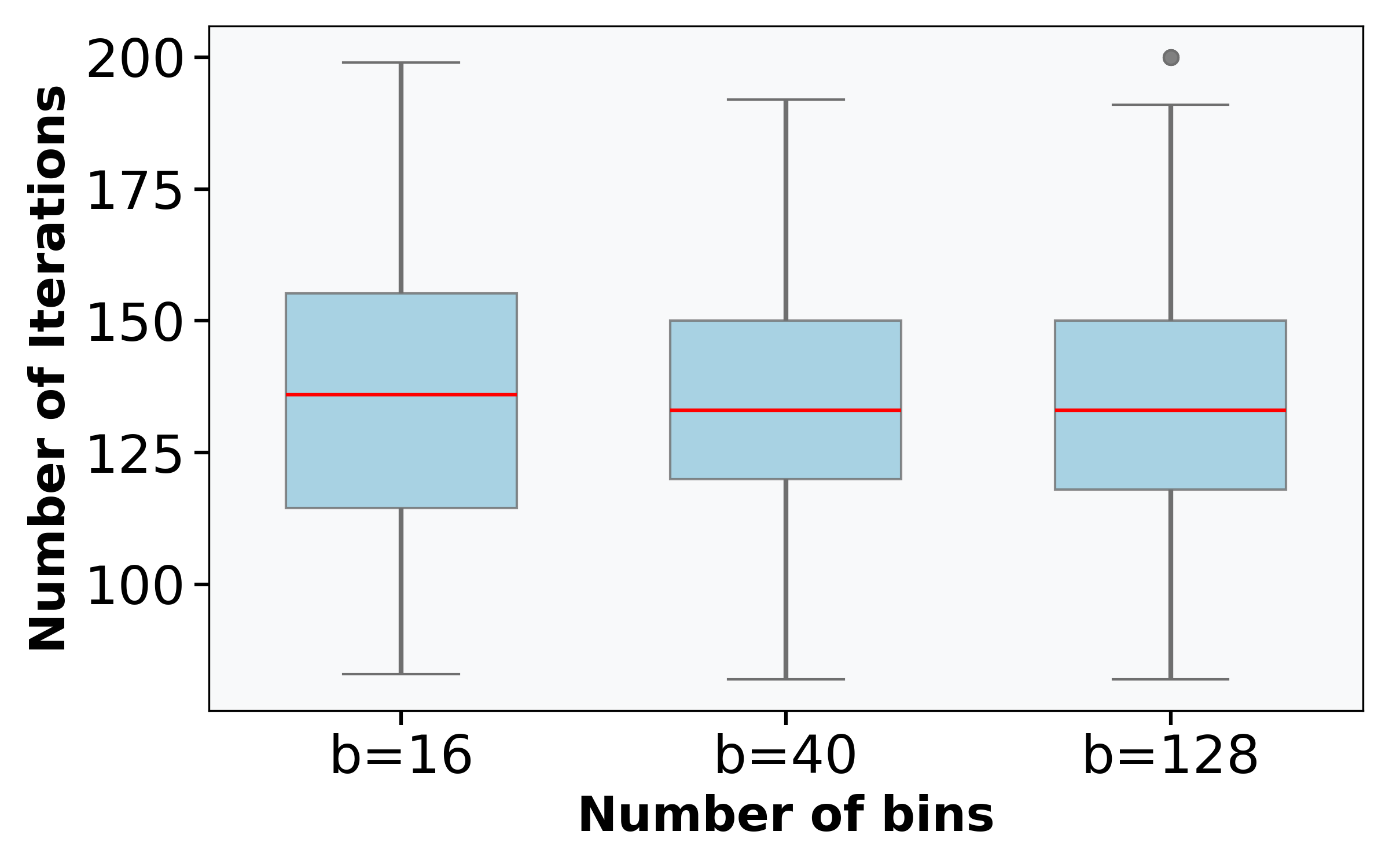}
        \captionsetup{justification=centering}
        \caption{Weak correlation \\($\beta=0.03$).}
        \label{fig:iterweak}
    \end{subfigure}
    \hfill
    \begin{subfigure}[b]{0.24\textwidth}
        \centering
        \includegraphics[width=\textwidth]{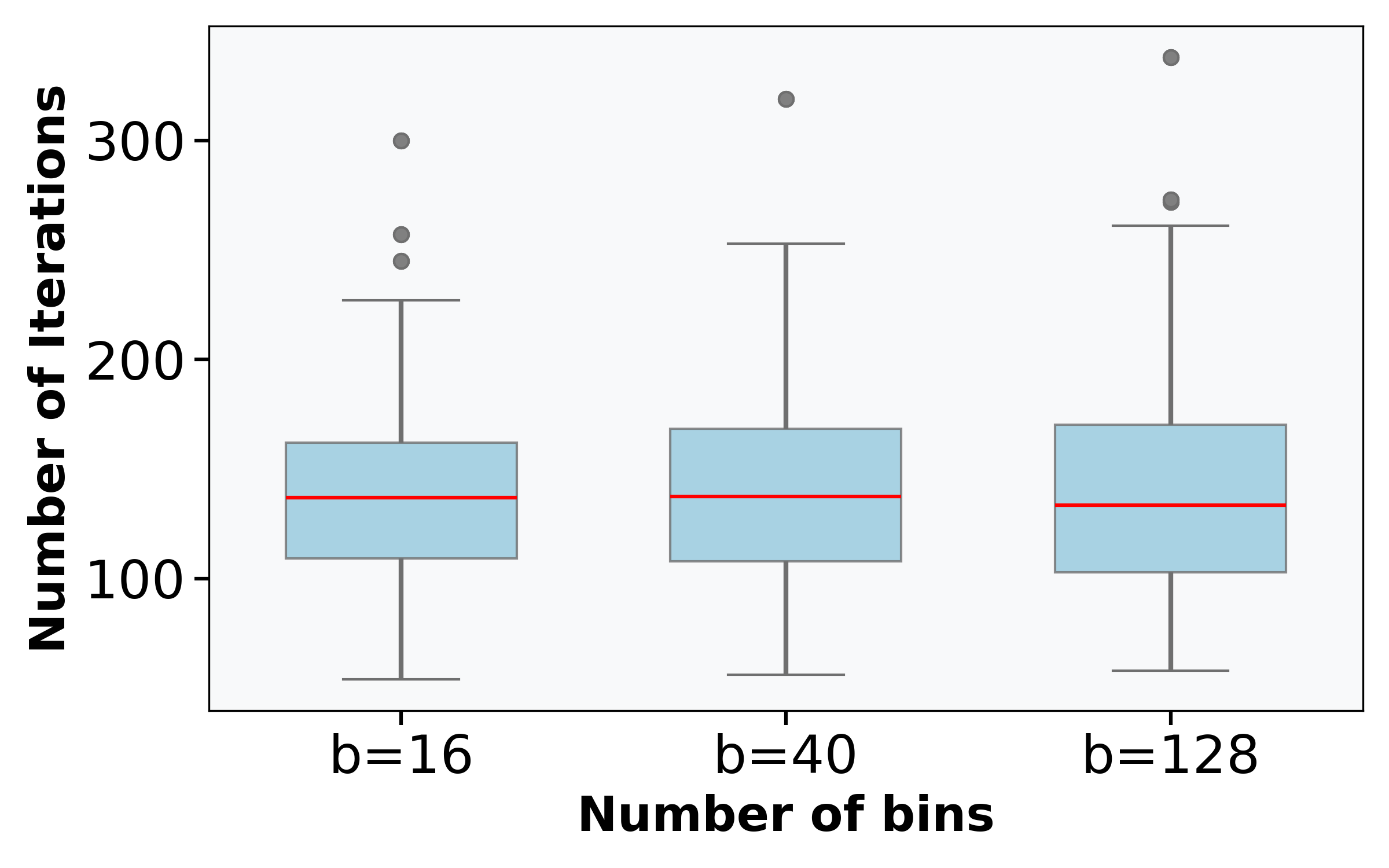}
        \captionsetup{justification=centering}
        \caption{Medium correlation \\($\beta=0.1$).}
        \label{fig:itermedium}
    \end{subfigure}
    \hfill
    \begin{subfigure}[b]{0.24\textwidth}
        \centering
        \includegraphics[width=\textwidth]{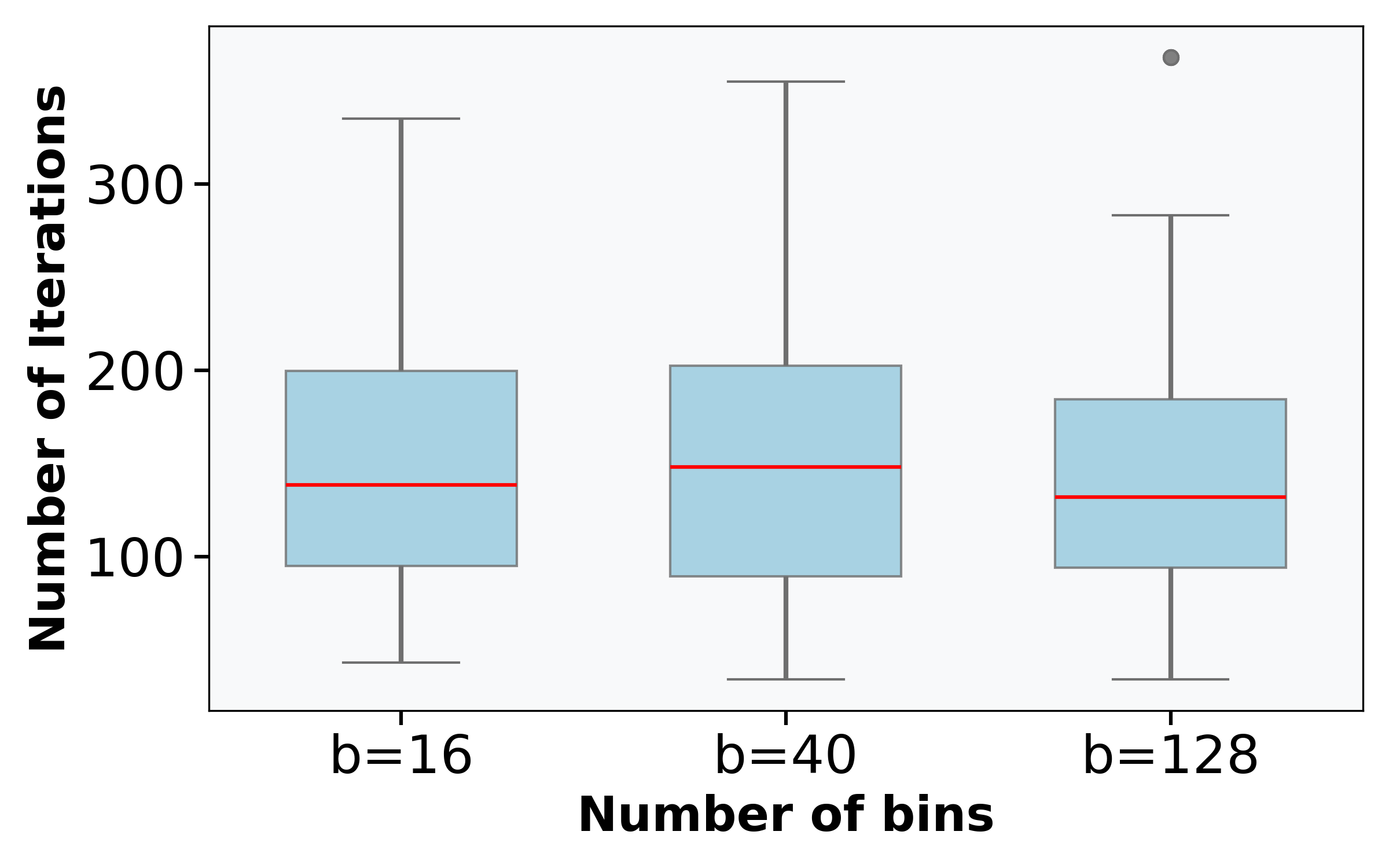}
        \captionsetup{justification=centering}
        \caption{Strong correlation \\($\beta=0.3$).}
        \label{fig:iterstrong}
    \end{subfigure}
    \caption{The number of iterations for MLE optimization over 100 replicas using the refined algorithm was evaluated with \(b = 16\), \(40\), and \(128\) bins across different correlation levels with problem size $51{,}076$, when $nu=0.5$.}
    \label{fig:iternbins}
\end{figure}

The numerical approximation accuracy of the \textsc{BesselK} function, as defined in Equation \eqref{eq:refined}, is significantly influenced by the discretization parameter $b$ (number of bins). Following the previous discovery, we vary the number of bins (\(b = 16, 40,\text{ and } 128\)) to assess its impact on the estimated parameters under weak, medium, and strong correlation levels when $\nu = 0.5$ and sample size $N = 51{,}076$, as illustrated in Figure~\ref{fig:mlenbins}. Additionally, we analyze the effect of the number of bins on the iteration count required for convergence across the three correlation levels, as shown in Figure~\ref{fig:iternbins}.

Although a smaller number of bins might introduce larger approximation errors given fixed integration bounds, this does not significantly affect the MLE procedure when $\nu$ is small. 
Through extensive numerical experiments (Figures \ref{fig:mle-combined}, \ref{fig:strong1-40}, \ref{fig:mlenbins}, and \ref{fig:iternbins}), we have empirically demonstrated that parameter estimation remains robust even with a reduced number of bins, supporting our hypothesis that accurate estimation can be achieved without requiring fine-grained discretization, although discretization should be chosen based on the specific application and desired accuracy level according to the aforementioned analysis. Overall, our approach suggests a practical balance between computational efficiency and numerical accuracy in evaluating the Mat\'{e}rn covariance function.


\subsection{Wind Speed Application}

\begin{figure}[!hbt]
    \centering
    \includegraphics[width=1.0\linewidth]{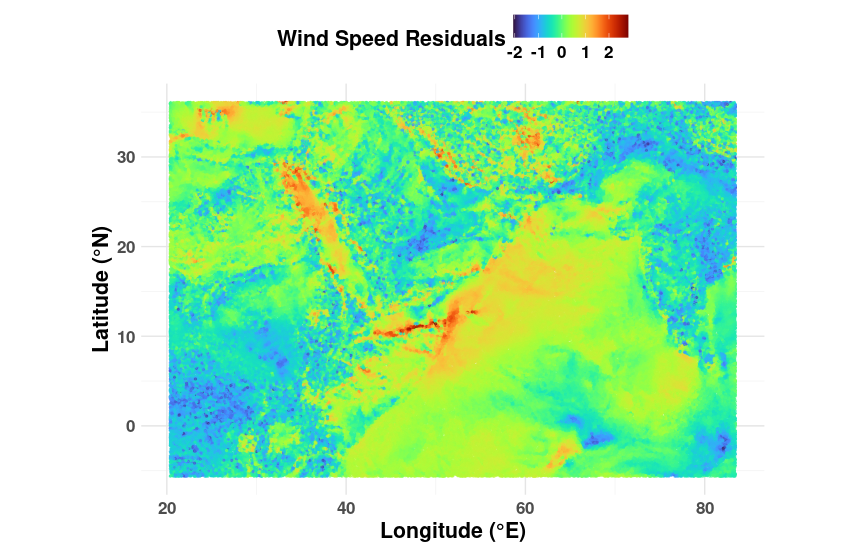}
    \caption{Residuals of a wind speed dataset of 1M locations in the Middle East region.}
    \label{fig:windres}
\end{figure}

For real data analysis in this study, we use a wind speed dataset of 1M locations generated using the WRF-ARW model for the Arabian Peninsula. The model has a horizontal grid resolution of \(5\) km and \(51\) vertical levels, extending to a maximum altitude of \(10\) hPa. The dataset spans a geographical region from \(20^\circ\)E to \(83^\circ\)E longitude and \(5^\circ\)S to \(36^\circ\)N latitude, covering \(37\) years of daily records. Each file contains hourly wind speed measurements across \(17\) atmospheric layers. Our analysis specifically focused on wind speed data from September 1, 2017, at 00:00 AM, examining measurements at 10 meters above ground level (layer 0). To address the skewed distribution of wind speeds, we plot the square root of wind speed against longitude and latitude, as shown in Figure~\ref{fig:windres}. 

Starting with an initial dataset of 1M locations, we randomly sampled $160{,}000$ locations for modeling and $25{,}000$ for testing. To improve numerical stability, the location coordinates are preprocessed through normalization. Given the length and width of a squared region $\ell_1, \ell_2$: (1) Compute the scaling factor $\ell := \max(\ell_1, \ell_2)$. (2) Rescale location $(x_0, y_0)$  to $(\{x_0 - \min(x_0)\}/\ell, \{y_0 - \min(y_0)\}/\ell)$. This transformation maps the spatial coordinates into a unit square $[0,1] \times [0,1]$, which helps mitigate numerical issues in subsequent computations.

Table~\ref{tab:comparison_table} presents the estimated parameters and the final log-likelihood values obtained by the GSL and the refined algorithms. The experiments were carried out on a 40-core Intel Cascade Lake CPU with $383$GB of memory and a $80$GB single A100 GPU. Both methods estimated nearly identical parameters, achieving nearly the same maximum log-likelihood value (llh), and produced the same mean square prediction error (MSPE).

However, using the GSL library required $596.61$ minutes of execution time, while the refined algorithm reduced this to $299.89$ minutes. For fairness, only the matrix generation was performed either by the CPU (GSL) or the GPU (refined algorithm), while all other operations were executed on the GPU.

\begin{table}[h!]
\caption{Comparison of GSL and refined algorithms on \( 160\text{K} \) random locations from the wind speed dataset.}
\label{tab:comparison_table}
\centering
\resizebox{\columnwidth}{!}{%
\begin{tabular}{|c|c|c|c|c|}
\hline
     & ($\sigma^2$, $\beta$, $\nu$) & llh    &MSPE       & Time\\ \
          & estimates & value    &       & (min) \\ \hline
GSL        & (2.505,      0.178,     0.426)         & 6984.660     &   0.037188    & 596.61       \\ \hline
Refined    & (2.510, 0.179, 0.426)    &  6984.598            &  0.037186    & 299.89       \\ \hline
\end{tabular}
}
\end{table}


\subsection{Performance Assessment}

For single-node performance evaluation, we use a 40-core Intel Cascade Lake CPU to run the GSL version and generate an entire covariance matrix within the \emph{ExaGeoStat} framework for varying numbers of locations. Additionally, we use 1 to 4 V100 (32 GB) or A100 (80 GB) GPUs on a single node to evaluate the implementation of the refined algorithm.

\begin{figure}[!hbt]
    \centering
    \includegraphics[width=0.45\textwidth]{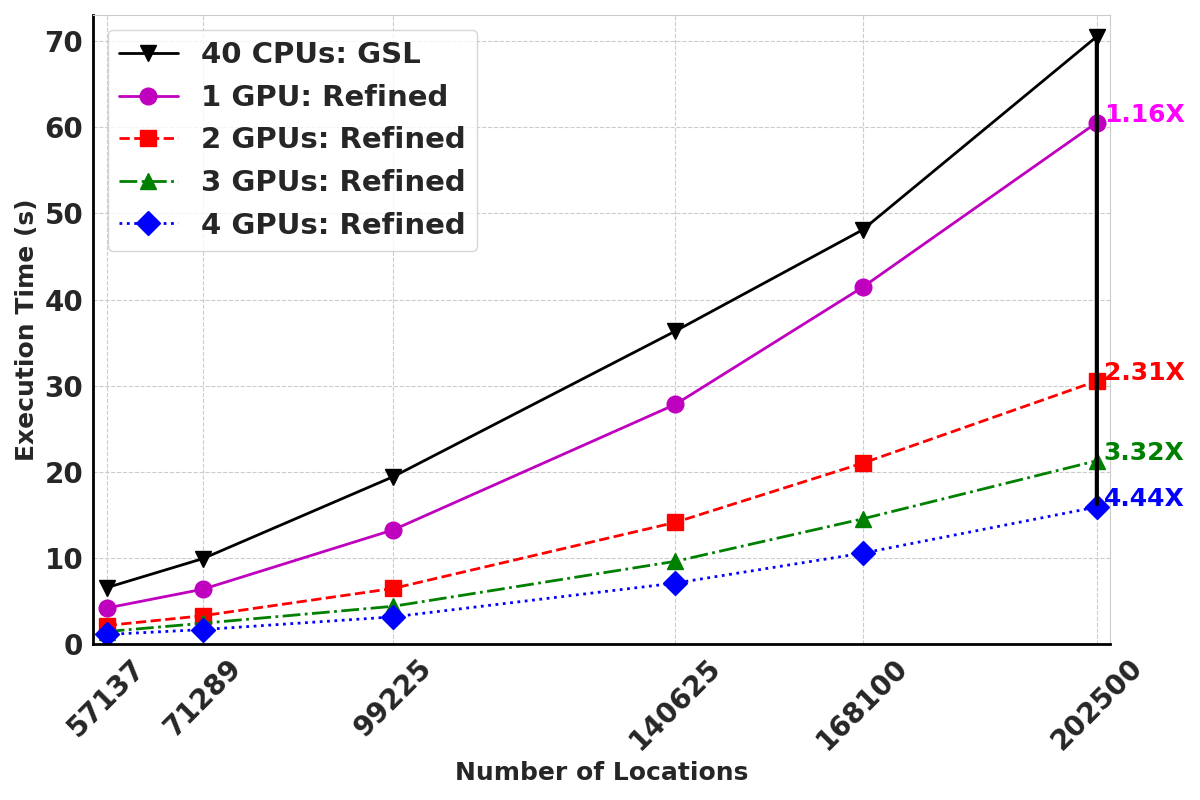}
    \caption{Execution time for generating an \( N \times N \) covariance matrix using GSL and the refined algorithm on a 40-core Intel Cascade Lake CPU and 1-4 NVIDIA V100 GPUs.}
    \label{fig:v100comparison}
\end{figure}

\begin{figure}[!hbt]
    \centering
    \includegraphics[width=0.45\textwidth]{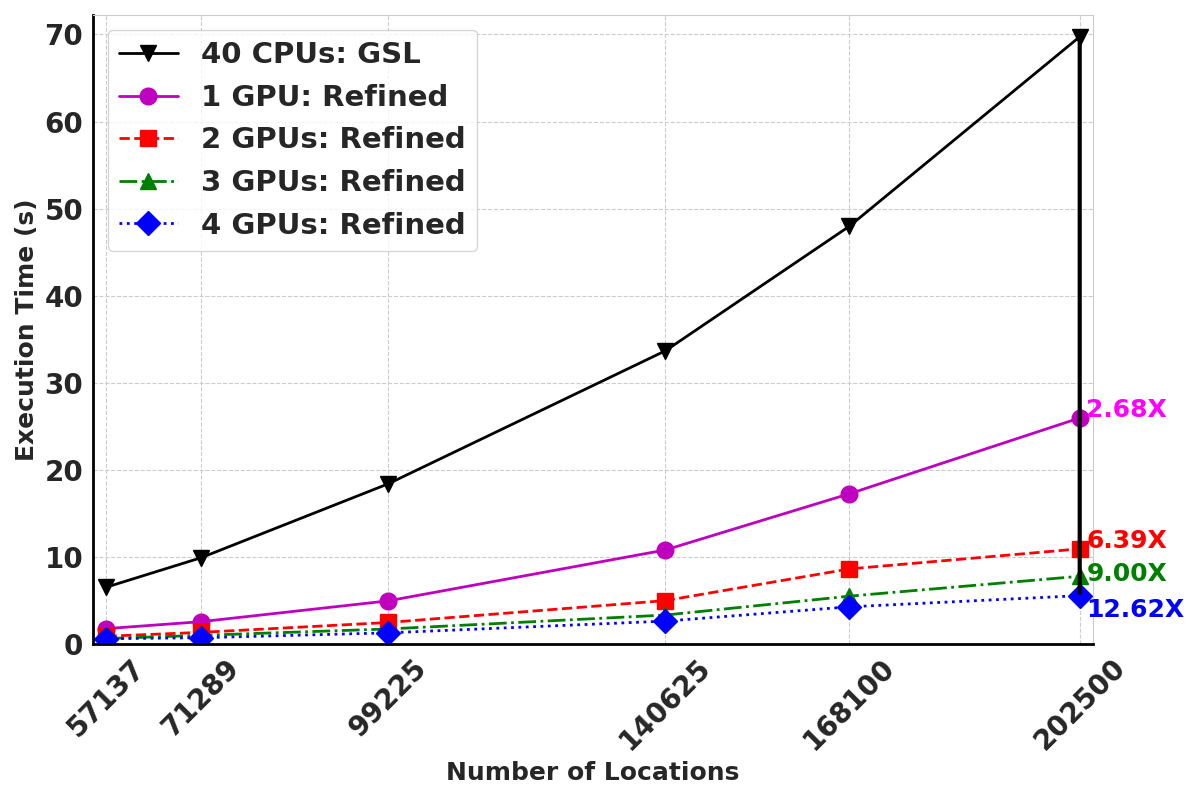}
    \caption{Mean execution time (s) for generating an \( N \times N \) covariance matrix using GSL and the refined algorithm on a 40-core Intel Cascade Lake CPU and 1-4 NVIDIA A100 GPUs.}
    \label{fig:a100comparison}
\end{figure}

In the single-GPU (V100) configuration (Figure~\ref{fig:v100comparison}), tests on datasets with a number of locations ranging from \(57{,}137\) to \(202{,}500\) demonstrated that the refined algorithm achieved a $1.16$X speedup compared to the CPU-only implementation. The performance improvements became increasingly pronounced with additional GPUs, reaching a $4.44$X speedup with four GPUs. For NVIDIA A100 GPUs, the single-GPU configuration (Figure~\ref{fig:a100comparison}), tested on datasets with a number of locations ranging from \(57{,}137\) to \(99{,}225\), showed a $2.68$X speedup over the CPU-only implementation. This improvement scaled significantly with additional GPUs, achieving a $12.62$X speedup with four GPUs. These results highlight the effectiveness of GPU acceleration with the refined algorithm for matrix generation, with benefits becoming more pronounced when leveraging multiple GPUs and larger datasets.

\begin{figure}[!hbt]
    \centering
    \includegraphics[width=0.9\linewidth]{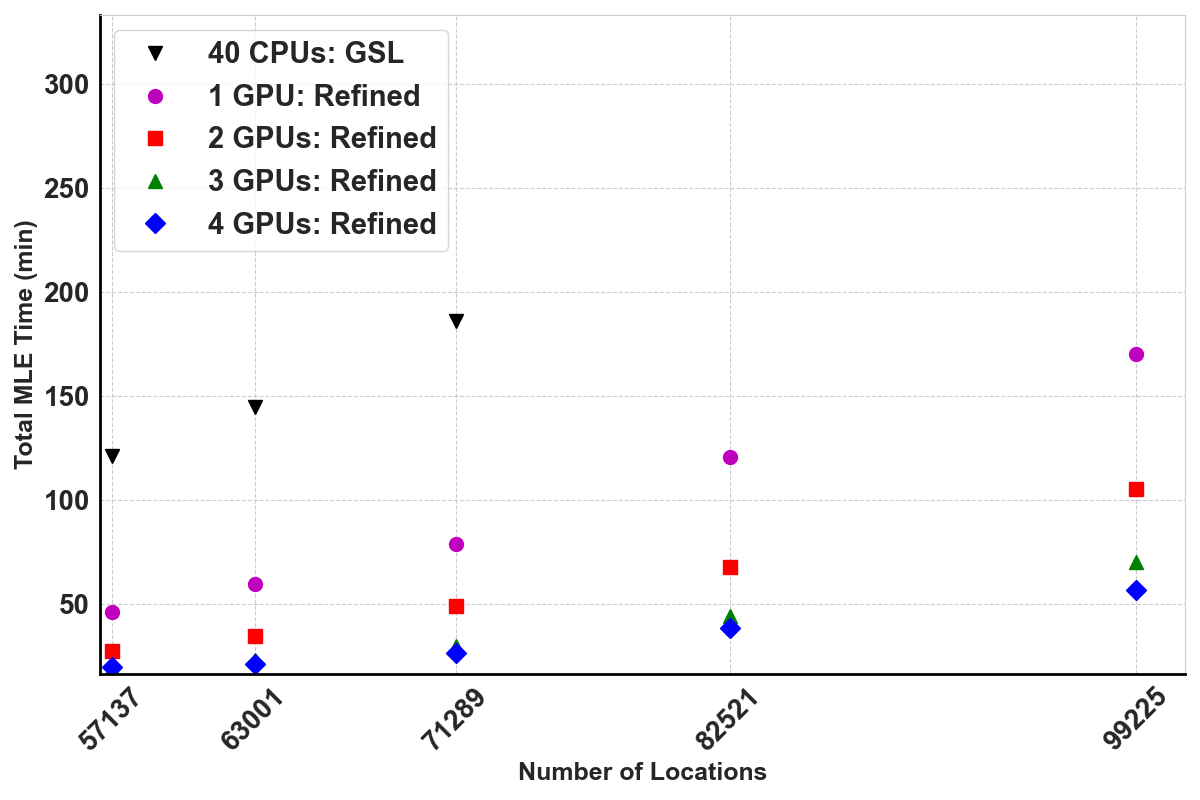}
    \caption{Comparison of overall MLE execution time between the GSL library and the refined algorithm on a V100 GPU across various problem sizes. The reported time accounts for multiple iterations of the log-likelihood function, including matrix generation and all associated linear algebra operations within the MLE process~\cite{abdulah2018exageostat}.}
    \label{fig:full_MLE_perf}
\end{figure}

Although the single-matrix generation time highlights the efficiency of the refined algorithm on GPUs compared to GSL, its impact becomes even more significant in operations like MLE. During MLE, the matrix generation function is invoked multiple times while optimizing the log-likelihood function until convergence, amplifying the benefits of the refined algorithm. Figure~\ref{fig:full_MLE_perf} illustrates the performance of executing the full MLE process on five different problem sizes, comparing the GSL implementation with the refined algorithm using up to four A100 GPUs.  Due to the long execution time of the GSL function, the full MLE process was not estimated for \(N = 82,521\) and \(N = 99,225\). For \(N = 71,289\), the complete MLE process took 186.21 minutes with GSL (CPU) and $78.87$, $49.36$, $30.01$, and $26.38$ minutes using the refined algorithm on 1 GPU, 2 GPUs, 3 GPUs, and 4 GPUs, respectively.

\begin{figure}[!hbt]
    \centering
    \includegraphics[width=0.9\linewidth]{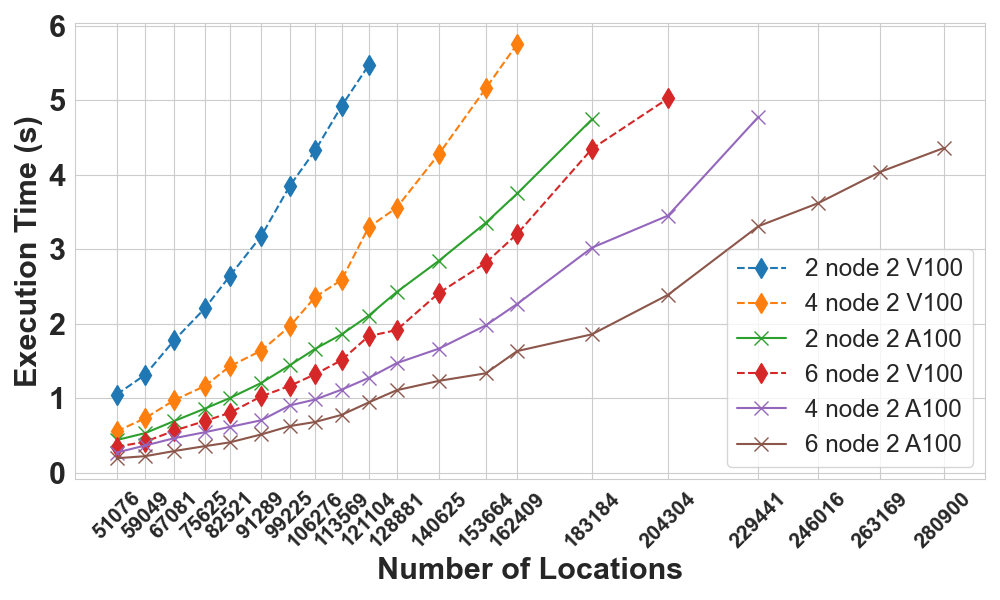}
    \caption{Mean matrix generation time varying node counts and GPU architectures (V100 and A100).}
    \label{fig:scaling}
\end{figure}

Figure~\ref{fig:scaling} shows the scalability of the refined algorithm across up to six nodes, each equipped with two V100/A100 GPUs. As available memory increases, GPUs can handle larger problem sizes and the performance scales almost linearly with more GPUs.

\section{Conclusion}
We present a highly efficient GPU-accelerated implementation of the modified Bessel function of the second kind (\textsc{BesselK}) using CUDA, effectively addressing critical computational bottlenecks in Gaussian processes and various other scientific applications. This approach can also be extended to other modeling frameworks, including non-Gaussian processes, where \textsc{BesselK} functions frequently arise. It integrates Temme’s series expansion for small input values and a refined version of Takekawa's integral-based approach for larger values, ensuring accuracy and computational efficiency across a reasonable parameter space.
Incorporating this optimized algorithm into the \emph{ExaGeoStat} framework demonstrates significant performance improvements in generating covariance matrices for spatial data modeling. The GPU-based implementation achieved substantial speedups compared to traditional CPU-based methods while maintaining high numerical accuracy, validated through synthetic datasets and real-world climate data. These improvements are particularly notable in large-scale applications that require massive matrix computations. Our work improves the computational capabilities of \textsc{BesselK} evaluations on GPUs and establishes the foundation for integrating such optimizations into other domains based on these functions. Future work includes extending the implementation to support the evaluation of derivatives of \textsc{BesselK}, enabling gradient-based optimization techniques, as well as transitioning from single-threaded to a multi-threaded evaluation of $K_{\nu}(x)$ to further improve performance.

\vspace{12pt}

\end{document}